%% file: main.tex
\documentclass[final,1p,times]{elsarticle}

 \usepackage{graphics}

\usepackage{amsmath}
\usepackage{amsthm}
\usepackage{xcolor}
\usepackage{amssymb}
\usepackage{amsfonts}
\usepackage{fancybox}
\usepackage{color}
\usepackage{bm}
\usepackage{cancel}

\usepackage{graphicx}
\usepackage{epstopdf}
\usepackage{algorithm}
\usepackage{algorithmic}

\usepackage{wrapfig}
\usepackage{caption}
\usepackage{graphicx,color}
\usepackage{subfigure}
\usepackage{bbm}
\usepackage{dsfont}
\usepackage{multirow}
\usepackage{amsmath,amsfonts,amssymb}
\DeclareMathOperator*{\argmin}{arg\,min}

\usepackage{geometry}
\usepackage{siunitx}

\newcommand{\bx}{\mathbf{x}}

\newcommand{\bu}{\mathbf{u}}

\newcommand{\bP}{\mathbf{P}}

\begin{document}
\begin{frontmatter}
\title{A spatially adaptive high-order meshless method for fluid-structure interactions}
\author[WISC]{Wei Hu}
\author[NAT]{Nathaniel Trask}
\author[TU]{Xiaozhe Hu}
\author[WISC]{Wenxiao Pan\corref{cor}}
\ead{wpan9@wisc.edu}
\cortext[cor]{Corresponding author}
\address[WISC]{Department of Mechanical Engineering, University of Wisconsin-Madison, Madison, WI 53706, USA}
\address[NAT]{Sandia National Laboratories, Albuquerque, NM 87185, USA\footnote{Sandia National Laboratories is a multimission laboratory managed and operated by National Technology and Engineering Solutions of Sandia, LLC.,
a wholly owned subsidiary of Honeywell International, Inc., for the U.S. Department of Energy’s National Nuclear Security Administration under contract DE-NA-0003525.}}
\address[TU]{Department of Mathematics, Tufts University, Medford, MA 02155, USA}

\begin{abstract} 

We present a scheme implementing an \textit{a posteriori} refinement strategy in the context of a high-order meshless method for problems involving point singularities and fluid-solid interfaces. The generalized moving least squares (GMLS) discretization used in this work has been previously demonstrated to provide high-order compatible discretization of the Stokes and Darcy problems, offering a high-fidelity simulation tool for problems with moving boundaries. The meshless nature of the discretization is particularly attractive for adaptive $h$-refinement, especially when resolving the near-field aspects of variables and point singularities governing lubrication effects in fluid-structure interactions. We demonstrate that the resulting spatially adaptive GMLS method is able to achieve optimal convergence in the presence of singularities for both the div-grad and  Stokes problems. Further, we present a series of simulations for flows of colloid suspensions, in which the refinement strategy efficiently achieved highly accurate solutions, particularly for colloids with complex geometries.

\end{abstract}


\end{frontmatter}

\input{introduction}

\input{method}

\input{results}
\input{conclusion}

\bibliographystyle{elsarticle-num}
\biboptions{numbers,sort&compress}
\bibliography{reference}

\end{document}

%% file: introduction.tex
\section{Introduction}\label{sec:Introduction}

Fluid-structure interactions (FSI) problems are ubiquitous in science and engineering applications. Examples include, but are not limited to, complex fluids \cite{ActiveColloid_ReviewAlexeev2016,ActiveColloid_Review2016b,ma2015electric,bueno2015interaction,ten2014gravitaxis,goto2015purely}, additive manufacturing \cite{randles2017computational,kamps2018design,sochol20163d}, and energy harvesting \cite{WindTurbineFSIIGA2012,EnergyWaveFSI2008,WindTurbineFSIIGA2016,EnergyWaveFSI2018}. Two traits shared by these problems are: \emph{i}) moving fluid-solid interfaces of complex geometries; and \emph{ii}) hydrodynamic interactions between many bodies. Thus, numerical methods for solving such problems must be able to efficiently track the moving interfaces and to accurately capture bidirectional fluid-solid hydrodynamic couplings, which can be challenging particularly when the solids display large displacements and rotations, and/or the moving solids have complex and feature-rich geometries. Moreover, in the zero Reynolds-number limit, solutions to the Stokes flow are well-known to exhibit singular pressures in the vicinity of boundaries with sharp corners and in narrow lubrication gaps occurring between solid bodies. In the flows of colloid suspensions, colloid kinematics of dense suspensions at zero-Reynolds number is dominated by the so-called lubrication effects. Thus, the accurate prediction of these flows depends on the accurate resolution of the singularities over a length scale potentially orders of magnitude smaller than the characteristic length of flow. Such suspension flows therefore provide a particularly interesting FSI application to apply adaptive schemes.

Meshless methods provide attractive and flexible discretization for FSI problems, due to their ability to circumvent costly management of deforming meshes. This is important not only in handling large deformations relevant to FSI, but also in handling the insertion and deletion of nodes in $h$-adaptive schemes without the need to preserve mesh quality. In this work, we consider the generalized moving least squares (GMLS), a recently developed meshless method that has been shown to provide stable high-order solution to the div-grad~\cite{Trask2017high} and steady Stokes \cite{Trask2018compatible} problems. GMLS provides a flexible optimization framework to obtain optimal recovery of linear target functionals over a desired reproduction space directly from scattered samples (GMLS particles), and is built upon a rigorous approximation theory~\cite{wendland2004scattered,mirzaei2012generalized}. In the div-grad case, stability can be obtained by constructing a staggered discretization analogous to mesh-based primal/dual discretizations; the flexibility of GMLS enables the $\epsilon$-ball graph of neighbor connectivity to be used as a surrogate for primal/dual meshes in, e.g., staggered finite volume methods. For the Stokes problem, GMLS enables the reconstruction of velocity over a space of divergence-free polynomials, which can be efficiently performed locally in a least-squares sense, contrary to mesh-based methods necessitating costly global construction of divergence-free shape functions. The spatial compatibility of these approaches in GMLS stands in contrast to our previous work using smoothed particle hydrodynamics (SPH) \cite{Pan_LLNS_2014,Pan_CBFSPH_2014,Pan_ISPH2_2017,Trask2015SPH,Pan_CMRSPH_2017,Pan_ARSPH_2019}, a different meshless method in which to date researchers have been unable to simultaneously achieve spatial compatibility and consistency. 

As with traditional mesh-based methods utilizing quasi-uniform triangulations, GMLS suffers from deteriorating accuracy in the vicinity of corner singularities and discontinuities in material properties. Additionally, successfully resolving lubrication gaps in Stokes problems while maintaining a uniform resolution imposes a restrictive computational cost. To this end, we propose the introduction of spatial adaptivity into the GMLS discretization.

An appropriate adaptivity criterion must be selected to ensure recovery of optimal convergence in the presence of singularities. Here, we chose to consider an error estimator. In the literature of adaptive finite element methods (AFEM) for solving second-order elliptic partial differential equations (PDEs), there are mainly two types of \textit{a posteriori} error estimators -- the residual-based~\cite{VerfurthR1996a,nochetto2009theory} and recovery-based~\cite{Zienkiewicz1992,RecoverError_Book2000}. A residual-based error estimator is derived from the variational formulation, and hence, strongly depends on the problem being solved~\cite{nochetto2009theory} while remaining inappropriate for the strong-form discretization used in GMLS. The recovery-based error estimator is solely based upon the computed solution and does not depend on the specific governing PDEs \cite{RecoverError_Book2000}. Thus, a recovery-based error estimator is more applicable to the current scheme than the residual type error estimators. Moreover, although the theoretical justification of the recovery-based error estimator is still under investigation~\cite{Xu.J;Zhang.Z2004,RecoverError_Zhang2005}, in practice, it has been shown to provide more accurate estimation of true errors when compared to the residual-based error estimators ~\cite{RecoverError_Zhang2004,Xu.J;Zhang.Z2004}. Therefore, we pursued a recovery-based error estimator as the criterion to direct spatial adaptivity in the GMLS discretization.

Given the adaptive criterion, we proposed a four-step adaptive algorithm: $\text{SOLVE}  \rightarrow \text{ESTIMATE}  \rightarrow \text{MARK}  \rightarrow \text{REFINE}$, to adaptively refine the spatial resolutions. In the \text{SOLVE} step, the governing equations are numerically solved using the GMLS discretization, and all of the variables are updated. In the \text{ESTIMATE} step, the recovery-based \textit{a posteriori} error estimator is evaluated at all GMLS particles, which measures the difference between the direct and recovered gradient. Here, the recovered gradient at a GMLS particle is defined as the average of the direct solutions of the gradient (e.g., velocity gradient) obtained on its neighbor particles. In the \text{MARK} step, the GMLS particles with larger recovery-based errors are marked for refining, following a strategy similar to the D{\"o}rfler's strategy used in AFEM \cite{dorfler1996convergent,stevenson2007optimality,nochetto2009theory}. Finally, in the \text{REFINE} step, the GMLS particles marked are refined by splitting them into smaller particles. At each time step, these four steps are iterated until the total estimated error of all GMLS particles is no more than a preset tolerance. 

The proposed spatially adaptive GMLS method was first validated and assessed by solving the div-grad problems with corner singularity or subject to discontinuous coefficients. By comparing with the analytical solutions, we examined whether the adaptive GMLS solutions recover the optimal, theoretical convergence orders and how accurate the proposed recovery-based error provides estimations of the true errors. For GMLS, only a truncation error analysis for differential operators exists \cite{mirzaei2012generalized}, and the lack of formal error analysis renders the notion of optimal convergence ill-defined. In this work, we informally refer to optimal convergence as matching the convergence observed in previous work for non-singular solutions and regular domains \cite{Trask2017high,Trask2018compatible}. Next, we employed the adaptive GMLS for solving Stokes flows subject to bidirectional fluid-solid couplings with the FSI applications of colloid suspensions. By comparing the numerical results with the analytical solutions and experimental data, we examined the accuracy of the proposed adaptive GMLS method in solving the FSI problems. By comparing with the GMLS solutions with uniform resolutions, we demonstrated the efficiency gains of the adaptive GMLS to achieve the same accuracy. 

This paper is organized as follows. Section \ref{sec:GoverningEquations} presents the governing equations to be solved for the div-grad and steady Stokes problems, respectively. We provide in Section \ref{sec:GMLSDiscre} a concise statement of the GMLS discretization employed in this work. Section \ref{sec:adaptive} explains the proposed recovery-based error estimator and algorithm for adaptive refinement. In Section \ref{sec:Simu_results}, we validated the adaptive GMLS through numerical tests. We solved two div-grad problems and four Stokes flows with applications to colloid suspensions -- the div-grad problem in an L-shaped domain, div-grad problem with rough coefficients, Wannier flow, dynamics of an L-shaped active colloid, dynamics of two square colloids under a shear flow, and collective dynamics of active asymmetric colloidal dimers. In this exercise, the numerical results were compared with the analytical solutions, GMLS solutions with uniform resolutions, or experimental data. Finally, we conclude and suggest directions of future work in Section \ref{sec:Conclu}.

%% file: method.tex

\section{Governing equations}\label{sec:GoverningEquations}
The proposed method was employed for solving both diffusion (div-grad) and Stokes problems. The computational domain can be represented as $\Omega = \Omega_f \cup\Gamma$, where $\Omega_f$ is the sub-domain occupied by the fluid, and $\Gamma$ denotes the boundary. 
 
\subsection{Div-grad problem}\label{subsec:Elliptic}
The div-grad boundary value problem is governed by: 
\begin{equation}
  \begin{cases}
    -\nabla\cdot (\kappa\nabla\phi)= f & \text{for } \bx \in \Omega_f\\
    \phi = \phi_{\Gamma}  & \text{for } \bx \in \Gamma\\
  \end{cases} \; ,
  \label{equ:Elliptic}
\end{equation}
where $\kappa \in \mathbf{L^2}(\Omega)$ describes a potentially discontinuous diffusion coefficient; all $\kappa$, $f$, and $\phi_{\Gamma}$ are given data. 
 
\subsection{Steady Stokes problem}\label{subsec:Stokes}
For the Stokes problem, we consider the coupled dynamics of freely suspended colloids and steady Stokes flow of incompressible fluid. Assume the domain $\Omega$ contains $N_c$ colloids, and each of them with a boundary $\Gamma_n$, $n=1,\dots,N_c$. Assuming finite separation between any two colloids, the boundary of the domain may be partitioned into the disjoint union $\Gamma = \Gamma_w \cup \Gamma_1 \cdots \cup \Gamma_{N_c} $, where $\Gamma_w$ denotes the wall boundaries. Each colloid has a position $\mathbf{X}_n$, an orientation $\mathbf{\Theta}_n$, and undergoes rigid-body kinematics prescribed by a translational velocity $\dot{\mathbf{X}}_n$ and rotational velocity $\dot{\mathbf{\Theta}}_n$. We refer to the collection of all colloids as a suspension. 
For a given $\left\{\mathbf{X}_n,\mathbf{\Theta}_n\right\}_{n=1,\dots,N_c}$, the geometrical configuration of the suspension (i.e., the set $\left\{\Gamma_n\right\}_{n=1,\dots,N_c}$) is prescribed, and the fluid flow is coupled to the colloid motion via the following steady Stokes problem, providing a boundary value problem for the fluid velocity and pressure:
\begin{equation}
  \begin{cases}
    \frac{\nabla p}{\rho} -\nu \nabla^2 \bu  = \mathbf{f} & \text{for } \bx \in \Omega_f \\
    \nabla \cdot \bu = 0  & \text{for } \bx \in \Omega_f \\
    \bu = \mathbf{w}  & \text{for } \bx \in \Gamma_w \\
    \bu = \dot{\mathbf{X}}_n + \dot{\mathbf{\Theta}}_n \times (\bx - \mathbf{X}_n)   & \text{for } \bx \in \Gamma_n,~n=1,\dots,N_c \; ,
  \end{cases}
  \label{equ:StokesEqn1}
\end{equation}
where $\nu$ is the kinematic viscosity of fluid; $\mathbf{f}$ denotes a body force such as gravity exerted on the fluid; $\mathbf{w}$ is the velocity of the wall boundaries, and $\mathbf{w}=\mathbf{0}$ for stationary walls. 

For a divergence-free velocity field, the vector identity $\nabla^2 \bu = -\nabla \times \nabla \times \bu$ holds. Thus, Eq. \eqref{equ:StokesEqn1} may be recast in the following equivalent form: 
\begin{equation}
  \begin{cases}
    \frac{\nabla p}{\rho} + \nu \nabla \times \nabla \times \bu  = \mathbf{f} & \text{for } \bx \in \Omega_f \\
         \frac{\nabla^2 p}{\rho} = \nabla \cdot \mathbf{f}    & \text{for } \bx \in \Omega\\
    \bu = \mathbf{w}  & \text{for } \bx \in \Gamma_w \\
    \bu = \dot{\mathbf{X}}_n + \dot{\mathbf{\Theta}}_n \times (\bx - \mathbf{X}_n)  & \text{for } \bx \in \Gamma_n, n=1,\dots,N_c \\
    \mathbf{n}\cdot \frac{\nabla p}{\rho} + \nu \mathbf{n} \cdot \nabla \times \nabla \times \bu= \mathbf{n} \cdot \mathbf{f}   & \text{for } \bx \in \Gamma \;.
    \end{cases}
  \label{equ:StokesEqn2}
\end{equation}
Here, $\mathbf{n}$ is the unit normal vector outward facing at boundary $\Gamma$. By recasting the Stokes equation \eqref{equ:StokesEqn1} in this equivalent form, the saddle-point structure of the Stokes operator can be avoided, and after discretization one obtains instead a system with elliptic matrices along the diagonal blocks that is more amenable to standard preconditioning techniques. This necessitates a discretization faithful to the divergence-free constraint. In the GMLS discretization, the velocity is directly constructed from an appropriate space of divergence-free vectors, as explained in Section \ref{sec:GMLSDiscre}. 

In general, the translational and angular dynamics of each colloid is governed by:
\begin{equation}
\begin{cases}
    M_n \ddot{\mathbf{X}}_n = \int_{\Gamma_n} \boldsymbol{\sigma} \cdot d\mathbf{A}  + \mathbf{F}_{e, n}\\
    I_n \ddot{\mathbf{\Theta}}_n = \int_{\Gamma_n}  \left( \bx - \mathbf{X}_n\right) \times (\boldsymbol{\sigma} \cdot d\mathbf{A}) + \mathbf{T}_{e, n}  \; ,
    \end{cases}
\label{equ:ForceTorque}
\end{equation}
where $M_n$ and $I_n$ denote the mass and moment of inertia of each colloid, respectively; $\boldsymbol{\sigma} = -p \mathbf{I} + {\nu}\left [\nabla \bu  + (\nabla \bu)^\intercal \right ]$ is the stress exerted by the fluid on colloids; $\mathbf{F}_{e, n}$ and $\mathbf{T}_{e, n}$ represent the external force and torque applied on each colloid, respectively. Assuming the inertia effect is negligible, which is valid for small colloids in a viscous fluid, the above dynamics of each colloid is subject to the force- and torque-free constraint; i.e., 
\begin{equation}
\begin{split}
    & \int_{\Gamma_n} \boldsymbol{\sigma} \cdot d\mathbf{A}  + \mathbf{F}_{e, n} = \mathbf{0} \\
    & \int_{\Gamma_n}  \left( \bx - \mathbf{X}_n\right) \times (\boldsymbol{\sigma} \cdot d\mathbf{A}) + \mathbf{T}_{e, n} = \mathbf{0}  \; .
\end{split}
\label{equ:ForceTorque_Free}
\end{equation}
Therefore, by solving Eqs. \eqref{equ:StokesEqn2} and \eqref{equ:ForceTorque_Free} concurrently as a monolithic system, we obtained a boundary value problem for the colloids' translational and angular velocities $\left\{\dot{\mathbf{X}}_n,\dot{\mathbf{\Theta}}_n\right\}_{n=1,\dots,N_c}$ as well as the fluid's velocity and pressure fields. Given $\left\{\dot{\mathbf{X}}_n,\dot{\mathbf{\Theta}}_n\right\}_{n=1,\dots,N_c}$, the colloids' positions and orientations $\left\{\mathbf{X}_n,\mathbf{\Theta}_n\right\}_{n=1,\dots,N_c}$ were numerically evolved by the 4th-order Runge-Kutta time-integration scheme \cite{butcher2016numerical}.

\section{GMLS discretization}\label{sec:GMLSDiscre}

To numerically solve the governing equations given in Section \ref{sec:GoverningEquations}, separate treatment was applied to the velocity $\bu$ and to the scalar fields $\phi$ and $p$ appearing in Eqs. \eqref{equ:Elliptic} and \eqref{equ:StokesEqn2}, respectively. While the velocity $\bu$ in the Stokes equations was handled via a GMLS approximation reconstructed over the space of divergence-free vector polynomials, $\phi$ in the diffusion equation and the pressure $p$ in the Stokes equations were approximated via a staggered GMLS discretization. In this section, we briefly review the basics of both the divergence-free \cite{Trask2018compatible} and staggered \cite{Trask2017high} approaches employed in this work. Regarding their derivation and analysis, we refer to the previous work \cite{Trask2017high,Trask2018compatible} for details. 

\subsection{GMLS approximation}\label{subsec:MLS}

The fluid domain $\Omega$ is discretized by a set of collocation points (called GMLS particles); another set of GMLS particles are distributed along the colloid and wall boundaries $\Gamma_w \cup \Gamma_1 \cdots \cup \Gamma_{N_c}$ for imposing the boundary conditions (BCs). We refer to these as interior and boundary particles, respectively. For a GMLS particle at $\bx_i$ and a given scalar function $\psi$ evaluated at its neighbor locations: $\psi_j = \psi(\bx_j)$, a polynomial $\psi_h(\bx)$ of order $m$ is sought to approximate $\psi$ and its derivatives $D^\alpha \psi$ at $\bx_i$. To this end, $\psi_h(\bx)=\bP^\intercal(\bx) \mathbf{c}^*$ with a polynomial basis $\bP(\bx)$ and coefficient vector $\mathbf{c}^*$ such that the following \emph{weighted} residual functional is minimized:
\begin{equation}\label{equ:MLS_min}
J(\bx_i) = \sum_{j \in {\cal{N}}_{\epsilon_i}} \left[ \psi_j - \bP_i^\intercal(\bx_j)\mathbf{c}_i \right]^2 W_{ij} \;.
\end{equation}
As such, for $\psi \in span(\bP(\bx))$, $\psi$ can be exactly reconstructed. From this polynomial reconstruction property, it follows that \emph{high-order accuracy} can be achieved via the GMLS approximation by taking large $m$, e.g., $m=4$ \cite{wendland2004scattered}. Following standard arguments for the minimization of a symmetric positive definite quadratic form, the solution of Eq. \eqref{equ:MLS_min} is given by:
\begin{equation}
\psi_h(\bx) = \bP_i^\intercal(\bx) \mathbf{c}^*_i ~~~~~~\text{with}~~~~\mathbf{c}^*_i= \left(\sum_{k \in {\cal{N}}_{\epsilon_i}} \bP_i(\bx_k) W_{ik} \bP_i^\intercal(\bx_k)\right)^{-1} \left( \sum_{j \in {\cal{N}}_{\epsilon_i}} \bP_i(\bx_j) W_{ij} \psi_j \right)\; . 
\label{equ:MLS_approx}
\end{equation}
To approximate derivatives of the underlying function, an arbitrary $\alpha^{th}$ order differential operator can be written as \cite{mirzaei2012generalized}:
\begin{equation}
D^\alpha \psi (\bx) \approx D^\alpha_h \psi_h (\bx) := (D^\alpha \bP (\bx))^\intercal \mathbf{c}^*\; .
\label{equ:diffuse_diff}
\end{equation}

In Eqs. \eqref{equ:MLS_min} and \eqref{equ:MLS_approx}, the weight function $W_{ij}=W(r_{ij})$, where $r_{ij}=\left \|\bx_i-\bx_j\right \|$; $W(r)=1-(\frac{r}{\epsilon})^4$ for $r<\epsilon$, otherwise, $W(r)\equiv 0$, with $\epsilon$ the compact support. Hence, it is only necessary to include GMLS particles within an $\epsilon$-neighborhood of the $i$th GMLS particle, i.e., $j \in {\cal{N}}_{\epsilon_i} = \left\{ \textbf{x}_j~ \text{s.t.}~ \left \|\textbf{x}_i - \textbf{x}_j\right \| <  \epsilon_i \right\}$. As a result, the approximation error in both the interpolant and derivatives of $\psi$ is governed by the choice of $\epsilon$. While a smaller $\epsilon$ yields higher accuracy and sparser linear operators, $\epsilon$ must be chosen sufficiently large to ensure unisolvency over the reconstruction space, and thus, a solution to the quadratic program \cite{wendland2004scattered}. In this work, $\epsilon_i = 4.5\Delta{x}_i$ with $\Delta{x}_i=V_i^{1/d}$, where $V_i$ is the volume occupied by particle $i$, and $d$ denotes the dimension, e.g., $d=2$ in 2D. For the adaptive strategy considered in this work, particles can have different $\Delta{x}$, and in turn, different $\epsilon$. We therefore require a definition of $W_{ij}$ whose support adapts to the local particle density in the vicinity of $\bx_i$. To ensure a symmetric weighting ($W_{ij}=W_{ji}$), we define the weight function as:
\begin{equation}
W_{ij} = \frac{W_{\epsilon_i}(r_{ij}) + W_{\epsilon_j}(r_{ij})}{2} \; .
\end{equation}

\subsection{Divergence-free GMLS reconstruction for the velocity $\bu$}\label{subsec:DivFreeU}
Based on the GMLS approximation discussed above, we seek a polynomial reconstruction $\bu_h(\bx)=(\bP^{div})^\intercal(\bx) \mathbf{c}^*$ to approximate $\bu$ and to discretize its gradient and curl-curl operators. To enforce compatibility with the divergence-free constraint for the velocity, we chose the polynomial basis $\bP^{div}(\bx)$ from the space of $m^{th}$ order divergence-free vector polynomials. 
For example, in 2D and for $m=2$, a divergence-free vector polynomial basis $\bP^{div}(\bx)$ at particle $i$ is given by:
\begin{equation}
\label{equ:divfreebasis}
\bP^{div}_i(\bx) =\left[ 
\binom{1}{0},
\binom{0}{1},
\binom{0}{-2\frac{x-x_i}{\epsilon_i}},
\binom{\frac{x-x_i}{\epsilon_i}}{-\frac{y-y_i}{\epsilon_i}},
\binom{2\frac{y-y_i}{\epsilon_i}}{0},
\binom{0}{-3(\frac{x-x_i}{\epsilon_i})^2},
\binom{(\frac{x-x_i}{\epsilon_i})^2}{-2\frac{x-x_i}{\epsilon_i}\frac{y-y_i}{\epsilon_i}},
\binom{2\frac{x-x_i}{\epsilon_i}\frac{y-y_i}{\epsilon_i}}{-(\frac{y-y_i}{\epsilon_i})^2},
\binom{3(\frac{y-y_i}{\epsilon_i})^2}{0}\right].
\end{equation}

The vectorial extension of Eqs. \eqref{equ:MLS_approx} and \eqref{equ:diffuse_diff}, providing expressions for the reconstruction, gradient, and curl-curl operators is given by \cite{Trask2018compatible}:
\begin{equation}
\begin{split}
&\bu_h(\bx_i) = (\bP^{div}_i)^\intercal(\bx_i) \mathbf{c}^*_i,  ~~~~~~\text{with}~~~~\mathbf{c}^*_i= \mathbf{M}_i^{-1} \sum_{j \in {\cal{N}}_{\epsilon_i}} (\bP_i^{div})^\intercal(\bx_j) \cdot \bu_j W_{ij} , ~~~\mathbf{M}_i = \sum_{j \in {\cal{N}}_{\epsilon_i}} \bP_i^{div}(\bx_j) W_{ij} (\bP_i^{div})^\intercal(\bx_j)\; , \\ 
\label{equ:MLS_approx_u}
&\nabla_h \bu_h (\bx_i) 
=  (\nabla\bP^{div}_i)^\intercal(\bx_i) \mathbf{c}^*_i \; , ~~~~~~
 (\nabla \times \nabla \times)_h \bu_h (\bx_i) 
=  (\nabla \times \nabla \times\bP^{div}_i)^\intercal(\bx_i) \mathbf{c}^*_i \; .
\end{split}
\end{equation}

The Dirichlet BC in Eq. \eqref{equ:StokesEqn2} for the velocity was then enforced by assigning the values of $\bu$ as specified in Eq. \eqref{equ:StokesEqn2} over the set of boundary GMLS particles.

\subsection{Staggered discretization of div-grad operator}\label{subsec:StaggeredP}

To ensure a compatible discretization of the div-grad operator for $\phi$ in the diffusion equation and for the pressure $p$ in the Stokes equations, a staggered GMLS discretization \cite{Trask2017high} was employed. Motivated by compatible mesh-based discretization on primal-dual grids \cite{CompatibleDis_Book2006,MimeticFD_JCP2014}, a local primal-dual complex was constructed for each GMLS particle via an $\epsilon$-neighborhood graph (virtual cell). In each \emph{virtual} cell, a set of primal edges can be built as: $\mathbf{E}_i = \{\bx_i-\bx_j ~|~ \bx_j \in {\cal{N}}_{\epsilon_i} \}$; and, each edge is associated with a midpoint $\bx_{ij}=\frac{\bx_i+\bx_j}{2}$ and a \emph{virtual} dual face at the midpoint and normal to the edge. The staggered GMLS discretization treats the div-grad operator as a composition of a GMLS divergence recovered from local virtual \emph{dual} faces together with a topological gradient over \emph{primal} edges, providing a generalization of familiar staggered finite difference methods to unstructured point clouds \cite{Trask2017high}. We succinctly describe below the staggered GMLS discretization for $\phi$, but it also applies to $p$. A detailed derivation and analysis of the discretization may be found in \cite{Trask2017high}.

Select as reconstruction basis $\bP$ the $\epsilon_i$-scaled Taylor monomials, which may be expressed in 2D as:
\begin{equation}
\label{equ:stagMLSbasisA}
\bP_i(\bx)=\left[\frac{1}{\alpha! \beta!}
\left(\frac{x-x_i}{\epsilon_i}\right)^{\alpha} \left(\frac{y-y_i}{\epsilon_i}\right)^{\beta} \right]_{|\alpha+\beta|\leq m}.
\end{equation}
Define the polynomial $q_i(\bx) =  \bP^{\intercal}_i(\bx)\mathbf{c}^*_i$, with coefficients determined by solving
\begin{equation}
\label{equ:stagMLSopt}
\mathbf{c}^*_i = 
\argmin\limits_{\mathbf{c}_i} 
\left\{ \sum_{j \in {\cal{N}}_{\epsilon_i}} \left[
\kappa_{ij}(\phi_j-\phi_i)  - \bP_i^\intercal(\bx_{ij}) \mathbf{c}_i  \right]^2 
W_{ij} \right\},
\end{equation}
where $\kappa_{ij} = \frac{\kappa(\bx_i) + \kappa(\bx_j)}{2}$. The optimal coefficient vector, gradient, and Laplacian are then provided by:
\begin{equation}
\label{equ:stagMLScoef}
\begin{split}
&\mathbf{c}^*_i =  \mathbf{M}_i^{-1} \sum_{j \in {\cal{N}}_{\epsilon_i}}\bP_i (\bx_{ij}) W_{ij} \kappa_{ij} (\phi_j-\phi_i), ~~~~~~\text{with}~~~ \mathbf{M}_i =  \sum_{j \in {\cal{N}}_{\epsilon_i}} \bP_i(\bx_{ij}) W_{ij} \bP_i^\intercal(\bx_{ij}) \; , \\
&\nabla_h \phi_h(\bx_i) = \frac12 (\nabla \bP_i)^\intercal(\bx_i)\mathbf{c}^*_i ~~~~~~~~~~~~~~~~~~~~~~~  \nabla^2_h \phi_h(\bx_i) = \frac14 (\nabla^2 \bP_i)^\intercal(\bx_i)\mathbf{c}^*_i\; .
\end{split}
\end{equation}
Evaluation of the gradient and Laplacian operators at interior GMLS particles follows from a direct evaluation of Eq. \eqref{equ:stagMLScoef}. For boundary GMLS particles, either a Dirichlet BC must be imposed in Eq. \eqref{equ:Elliptic}, or an inhomogeneous Neumann BC must be handled in Eq. \eqref{equ:StokesEqn2}. The Dirichlet BC was imposed by simply setting: $\phi_i = \phi_\Gamma (\bx_i)$. To impose an inhomogeneous Neumann BC of the form $\partial_{\mathbf{n}}|_{\Gamma} = g$, we added a single equality constraint to the quadratic program Eq. \eqref{equ:stagMLSopt} to enforce
\begin{equation}
    \mathbf{n}_i \cdot [\frac12 (\nabla  \bP_i)^\intercal(\bx_i)\mathbf{c}^*_i] = g_i,
\end{equation}
where $\mathbf{n}_i$ is the unit outward facing normal at boundary particle $i$. Further details regarding the imposition of Neumann BCs may be found in \cite{Trask2017high}.

\subsection{Approximation of integrals at colloid boundaries}\label{subsec:approx_int}

The integral of the stress tensor $\boldsymbol{\sigma}$ in Eq. \eqref{equ:ForceTorque_Free} requires the construction of a meshless quadrature rule. Consider the collection of boundary particles $\bx_i \in \Gamma_n$. We assumed the boundary may be partitioned into a disjoint collection of planar/curved faces $\Gamma_{ni}$. 
For each $\Gamma_{ni}$, we associated a unit outward facing normal $\mathbf{n}_i$ and measure of face area $\mathcal{A}_i$.
A composite quadrature rule for the integral may then be prescribed as: 
\begin{equation}
\label{equ:int_stress}
\int_{\Gamma_n} \boldsymbol{\sigma} \cdot d\mathbf{A} = \sum_{i \in \Gamma_n} (\boldsymbol{\sigma}_i \cdot \mathbf{n}_i) \mathcal{A}_i \; .
\end{equation}
The stress tensor $\boldsymbol{\sigma} = -p \mathbf{I} + {\nu}\left [\nabla \bu  + (\nabla \bu)^\intercal \right ]$ must be reconstructed at $\bx_i$, where $p$ is already available, and it only remains to approximate the gradient of the velocity $\nabla\bu$. Following the formulations in Section \ref{subsec:DivFreeU}, we approximated $\nabla\bu$ at each boundary particle. 
The integral was then approximated as:
\begin{equation}
\label{equ:int_stress_quadrature}
\int_{\Gamma_n} \boldsymbol{\sigma} \cdot d\mathbf{A} = \sum_{i \in \Gamma_n} \left(p_{i} \mathbf{I} + \nu \left[\nabla_h \bu_{h,i}+(\nabla_h \bu_{h, i})^\intercal\right]\right) \cdot \mathbf{n}_i \mathcal{A}_i,
\end{equation}
where $\nabla_h \bu_{h,i} = \nabla_h \bu_h(\bx_i)=(\nabla \bP^{div}_i)^\intercal(\bx_i) \mathbf{c}_i^* $.
For the purposes of this work, this quadrature rule sufficed to obtain the reported high-order convergence results, while the interested reader may consider referring to \cite{Trask2018compatible} for a discussion of higher-order quadrature rules.

\section{Algorithm for adaptive refinement}\label{sec:adaptive}

In this section, we introduce the algorithm for achieving adaptive refinement in the GMLS discretization. The algorithm relies on a recovery-based \textit{a posteriori} error estimator as the adaptive criterion. 

\subsection{Recovery-based error estimator}\label{subsubsec:Err_Cri}
For the div-grad problem, the error estimator was evaluated for $\phi$; for the Stokes problem, the errors were estimated  based on the velocity $\bu$. We describe below the proposed recovery-based error estimator for $\bu$, but it also applies to $\phi$.

Assuming the \emph{exact} solution of the velocity gradient $\nabla\bu$ and the \emph{direct} solution $\nabla_h\bu_h$, the \emph{true} error of the GMLS approximation at particle $i$ is:
\begin{equation}
\label{equ:true_error}
\| \mathbf{e}^t_i \|^2 = \sum_{j\in {\cal{N}}_{\epsilon_i}}\| \nabla \bu_j - \nabla_h \bu_{h, i \rightarrow  j} \| ^2.
\end{equation}
Here, the energy norm $\| \cdot \|^2$ is defined as the element-wise inner product; i.e., $\| \mathbf{e} \|^2 = \mathbf{e}: \mathbf{e} = e^{mn} e^{mn}$, where the Einstein summation convention is adopted. 

\begin{figure}[htbp]
	\centering
	\includegraphics[width=5.0in]{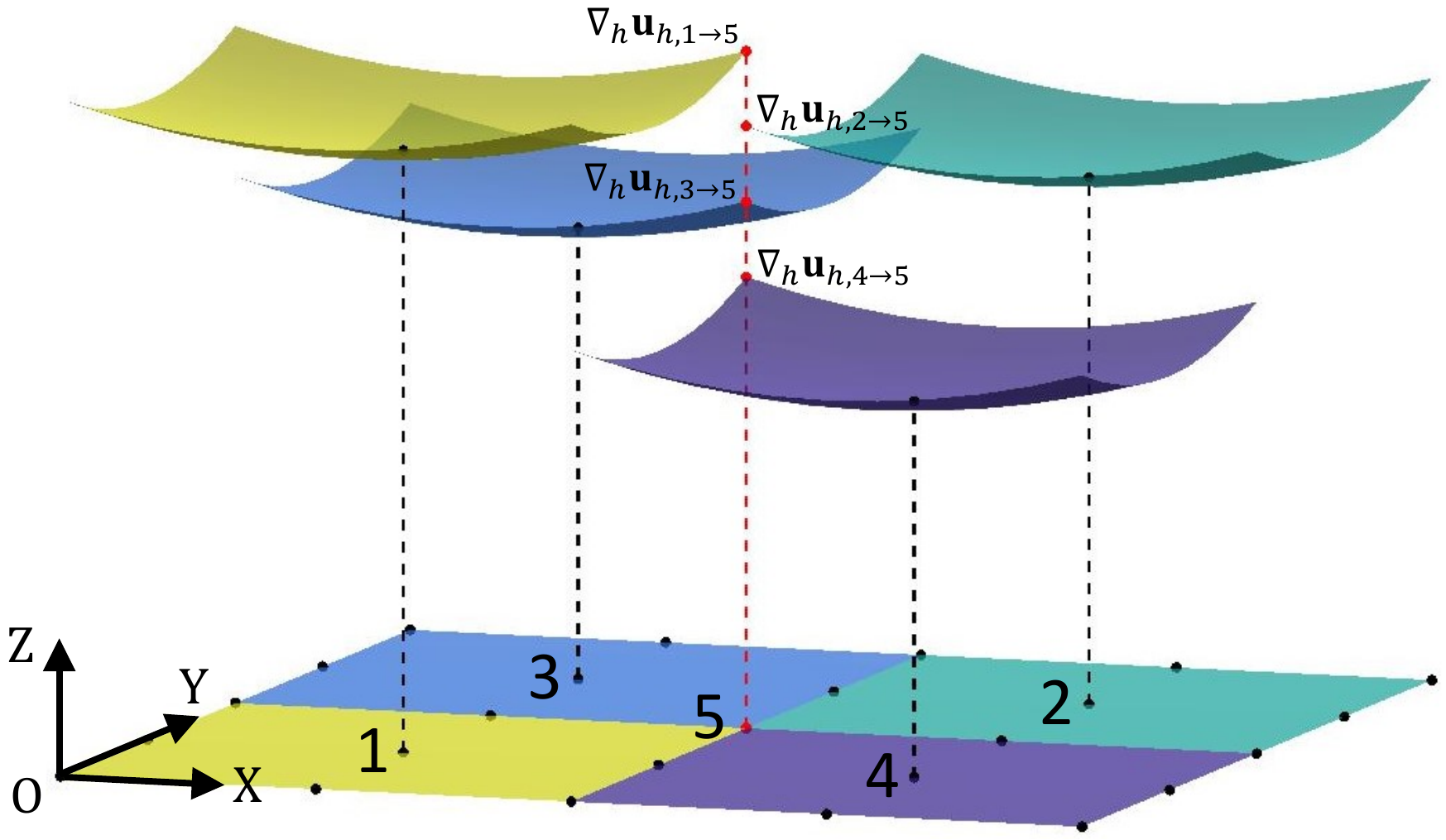}
	\caption{Schematic of the \emph{direct} solution $\nabla_{h} \bu_{h, i\rightarrow j}$. The curved surfaces with different colors represent  $\nabla_h\bu_h= (\nabla \bP^{div}_i)^\intercal(\bx) \mathbf{c}_i^*$ reconstructed at particles $i=$ 1, 2, 3, and 4, respectively. Particle 5 is a neighbor particle shared by particles 1, 2, 3, and 4. Evaluating $\nabla_{h} \bu_{h, i\rightarrow j}$ at particle 5 reads four different values: $\nabla_{h} \bu_{h, 1\rightarrow 5}$, $\nabla_{h} \bu_{h, 2\rightarrow 5}$, $\nabla_{h} \bu_{h, 3\rightarrow 5}$, and $\nabla_{h} \bu_{h, 4\rightarrow 5}$.} 
    \label{fig:Recovery_error}
\end{figure}

The direct solution $\nabla_{h} \bu_{h, i\rightarrow j}$ in Eq. \eqref{equ:true_error} was computed by reconstructing the velocity gradient at particle $i$ and evaluating it at particle $j$; i.e., 
\begin{equation}
\nabla_{h} \bu_{h, i\rightarrow j} = (\nabla \bP^{div}_i)^\intercal(\bx_j) \mathbf{c}_i^* .
\end{equation}
Thus, if $j$ was considered the neighbor of a different particle $i'$, although still evaluated at particle $j$, $\nabla_{h} \bu_{h, i' \rightarrow j}$ would be different. The definition of $\nabla_{h} \bu_{h, i\rightarrow j}$ is further illustrated in Figure \ref{fig:Recovery_error}, where the curved surfaces with different colors represent the velocity gradient reconstructed at different particles: $\nabla_h\bu_h = (\nabla \bP^{div}_i)^\intercal(\bx) \mathbf{c}_i^*$, $i=$ 1, 2, 3, and 4. Particle 5 is a neighbor particle shared by particles 1, 2, 3, and 4. Evaluating $\nabla_{h} \bu_{h, i\rightarrow j}$ at particle 5 ($j=5$) reads four different values. 

The true error as defined in Eq. \eqref{equ:true_error} cannot always be evaluated due to a lack of the exact solution in practical applications. Instead, the \emph{recovered} error, which measures the difference between the direct and recovered velocity gradient \cite{ainsworth2000posteriori}, can be practically determined in any simulation. In the GMLS, this recovered error can be evaluated locally on each GMLS particle at every time step as:
\begin{equation}
\label{equ:recover_error}
\| \mathbf{e}^r_i \|^2 =  \sum_{j\in {\cal{N}}_{\epsilon_i}}\| \mathbf{R}[\nabla\bu]_j - \nabla_{h} \bu_{h, i \rightarrow j}  \| ^2.
\end{equation}
Here, $\mathbf{R}[\nabla\bu]$ represents the recovered velocity gradient, and we proposed the following definition for it: 
\begin{equation}
\label{equ:recover_grad}
\mathbf{R}[\nabla\bu]_j = \frac{1}{N_j}\sum_{k\in {\cal{N}}_{\epsilon_j}} \nabla_{h} \bu_{h, k \rightarrow  j} = \frac{1}{N_j}\sum_{k\in {\cal{N}}_{\epsilon_j}} (\nabla \bP^{div}_k)^\intercal(\bx_j)
\mathbf{c}_k^*,
\end{equation}
where $k$ and $N_j$ denote the index and total number of the neighbor particles, respectively, of particle $j$. Thus, the recovered velocity gradient at particle 5 in Figure \ref{fig:Recovery_error} can be defined as: $\mathbf{R}[\nabla\bu]_5= \frac{1}{4}(\nabla_{h} \bu_{h, 1\rightarrow 5}+\nabla_{h} \bu_{h, 2\rightarrow 5}+\nabla_{h} \bu_{h, 3\rightarrow 5}+\nabla_{h} \bu_{h, 4\rightarrow 5})$, assuming it has only four neighbors.

Following the AFEM \cite{RecoverError_Zhang2004,RecoverError_Zhang2005}, being an appropriate adaptive criterion, the recovered error should have the same convergence order as the true
error; i.e., the \textit{a posteriori} error estimator is reliable and robust. In Section \ref{sec:Simu_results}, we numerically assessed and validated the proposed recovery-based error estimator given by Eqs. \eqref{equ:recover_error} and \eqref{equ:recover_grad}. 

\subsection{Adaptive algorithm}\label{subsubsec:Adapt_Algo}  

Given the recovery-based \textit{a posteriori} error estimator as the adaptive criterion, we introduce the following adaptive algorithm. At each time step, starting from a \emph{uniform}, \emph{coarse} resolution, the spatial resolution was adaptively refined following a four-step iteration procedure: 

\begin{equation}\label{def:adap-iter}
\textbf{SOLVE}  \rightarrow \textbf{ESTIMATE}  \rightarrow \textbf{MARK}  \rightarrow \textbf{REFINE}
\end{equation}

In the \textbf{SOLVE} step, the governing equations, Eq. \eqref{equ:Elliptic} for the div-grad problem or Eqs. \eqref{equ:StokesEqn2} and \eqref{equ:ForceTorque_Free} for the Stokes problem, were numerically solved by GMLS, and all of the variables were updated. In particular, to efficiently solve the resulting linear system after the GMLS discretization, we employed different linear solvers for different equations. Specifically, for Eq. \eqref{equ:Elliptic}, since it is a Poisson problem, we used the algebraic multigrid (AMG) method. For Eqs. \eqref{equ:StokesEqn2} and \eqref{equ:ForceTorque_Free}, the resulting linear system has the following block structure:
\begin{equation}\label{eqn:2by2-mat}
\begin{bmatrix}
\mathbf{K} & \mathbf{G} \\
\mathbf{B} & \mathbf{L}
\end{bmatrix}
\begin{bmatrix}
\tilde{\mathbf{u}} \\
p
\end{bmatrix}
 = 
 \begin{bmatrix}
 \mathbf{f}_{tot} \\
 g
 \end{bmatrix}\; .
\end{equation}
Here, we combined the velocities at all interior and boundary GMLS particles and colloids' translational and angular velocities into the vector $\tilde{\mathbf{u}}$. $\mathbf{K}$ contains contributions from the curl-curl ($\nabla \times \nabla \times$) operator in Eq. \eqref{equ:StokesEqn2} and the velocity gradient in the stress $\boldsymbol{\sigma}$ in the force- and torque-free constraint \eqref{equ:ForceTorque_Free}. $\mathbf{G}$ encompasses the $\frac{1}{\rho}\nabla$ operator in Eq. \eqref{equ:StokesEqn2} and the contribution from the pressure in the stress $\boldsymbol{\sigma}$ in the force- and torque-free constraint \eqref{equ:ForceTorque_Free}. $\mathbf{L}$ corresponds to the Laplacian operator ($\nabla^2$) in Eq. \eqref{equ:StokesEqn2}. $\mathbf{B}$ is non-zero only at the boundaries and contains the contribution from the $\nu \mathbf{n} \cdot \nabla \times \nabla \times$ operator in the inhomogeneous Neumann BC in Eq. \eqref{equ:StokesEqn2}. On the right-hand side of Eq. \eqref{eqn:2by2-mat}, $\mathbf{f}_{tot}$ combines the body force $\mathbf{f}$, $\mathbf{F}_{e,n}$ and $\mathbf{T}_{e,n}$ applied on each colloid, and the velocity $\mathbf{w}$ of the wall boundaries; $g$ contains $\nabla\cdot\mathbf{f}$ and $\mathbf{n}\cdot\mathbf{f}$ in Eq. \eqref{equ:StokesEqn2}. Following the previous work~\cite{Trask2018compatible}, we used the preconditioned general minimal residual (PGMRes) method to solve the above linear system with a block triangular preconditioner given as follows:
\begin{equation}\label{eqn:prec}
\begin{bmatrix}
\mathbf{S}  & \mathbf{G} \\
\mathbf{0} & \mathbf{L}
\end{bmatrix}^{-1},
\end{equation}
where $\mathbf{S} = \mathbf{K} - \mathbf{G} \, \text{diag}(\mathbf{L})^{-1} \, \mathbf{B}$ is an approximate Schur complement. To employ this preconditioner, both $\mathbf{S}$ and $\mathbf{L}$ needed to be inverted. In our implementation, we used the AMG method for the inversion. Due to the singularity of the solutions and adaptive refinement, standard relaxation schemes, such as Gauss-Seidel smoother, were not effective in the AMG. Thus, we applied a drop-tolerance type incomplete LU (ILU) smoother~\cite{Saad2003a} on the finest level of the AMG to improve the robustness of AMG both for solving the linear system of the div-grad problem and inverting the matrices in the block preconditioner \eqref{eqn:prec} for the Stokes problem. 

In the \textbf{ESTIMATE} step, the \textit{a posteriori} recovered error was firstly estimated locally at each GMLS particle based on Eqs. \eqref{equ:recover_error} and \eqref{equ:recover_grad}. Next, the particles were ranked according to the magnitude of the estimated recovered error. Finally, the total error of all GMLS particles: $E_{tot}=\sum\limits_i \|\mathbf{e}^r_i\|^2 V_i$, was evaluated.

In the \textbf{MARK} step, a portion of GMLS particles were selected and marked for refining according to the local, recovered error estimated in the ESTIMATE step. The particles with larger estimated errors are the potential candidates for refinement. In practice, the top $N_t$ particles ranked in the ESTIMATE step with a total error of $\alpha_s E_{tot}$ were marked. This MARK strategy is similar to the D{\"o}rfler's strategy used in the AFEM \cite{dorfler1996convergent,stevenson2007optimality,nochetto2009theory}. Here, $\alpha_s \in (0, 1)\times 100\%$ is the preset marking percentage. The value of $\alpha_s$ should be chosen depending on the specific problem and according to the distribution of the estimated recovered errors. A higher marking percentage is recommended if the estimated errors vary by orders across different regions of the domain. Otherwise, a smaller $\alpha_s$ is sufficient. Note that a larger $\alpha_s$ calls for more particles marked and refined, thereby a more expensive simulation. 

\begin{figure}[htbp]
	\centering
	\includegraphics[width=6.0in]{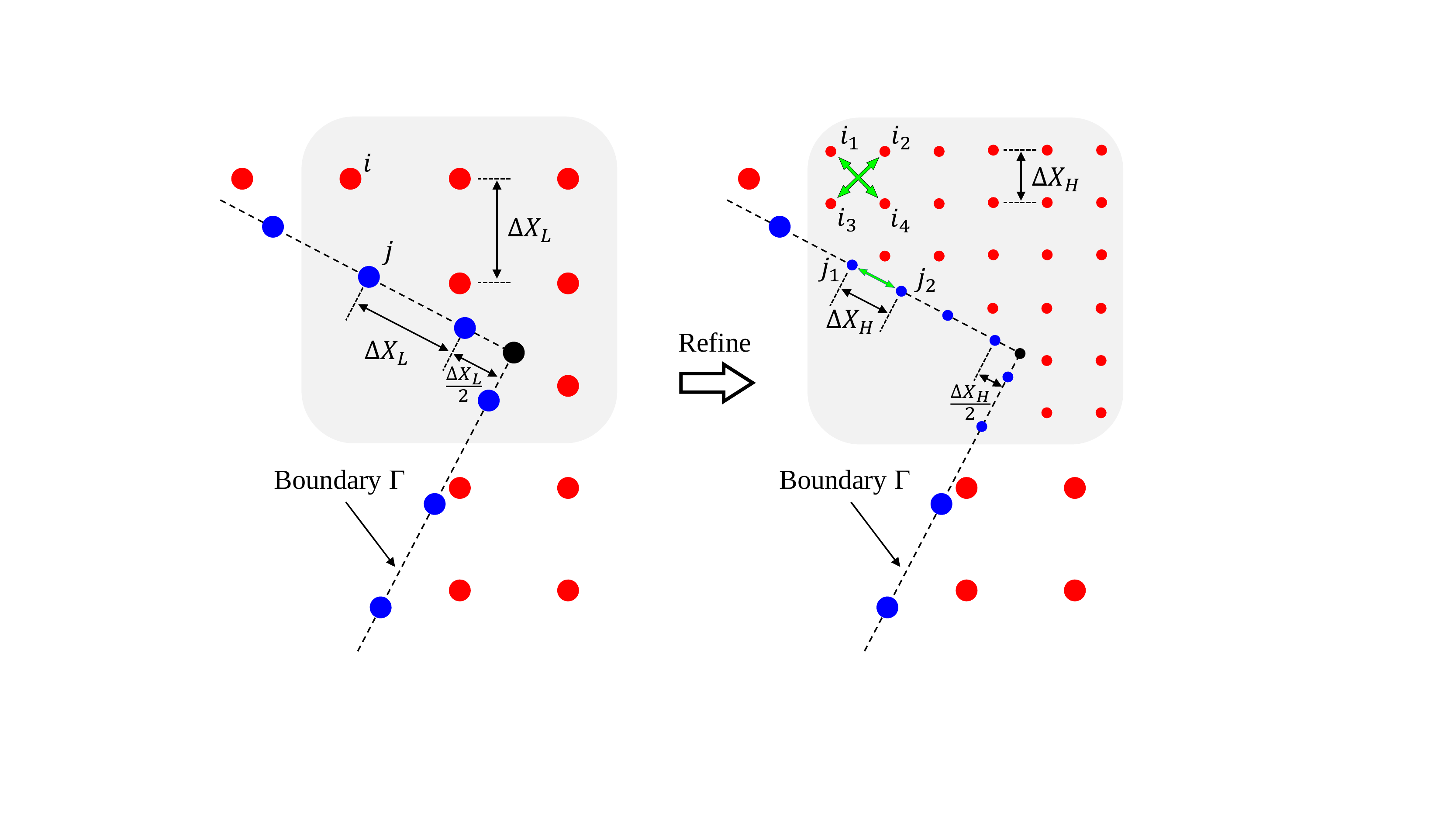}
	\caption{Schematic of the \textbf{REFINE} step. The GMLS particles in the gray area are marked for refinement, where each red, interior particle is split into 4 small particles with the spacing $\Delta X_H$, e.g., particle $i$ is split into $i_1$, $i_2$, $i_3$, and $i_4$; each blue, boundary particle is split into 2 small particles with the spacing $\Delta X_H$, e.g., particle $j$ is split into $j_1$ and $j_2$. The spacing length: $\Delta X_H = \Delta X_L/2$. In the presence of a sharp corner, there's always a GMLS particle (black) put right at the corner.} 
    \label{fig:Split-Process}
\end{figure}

In the \textbf{REFINE} step, the GMLS particles marked in the MARK step were refined. In particular, a marked interior GMLS particle was split into 4 small particles (in 2D); a marked boundary particle was split into 2 small particles, as illustrated in Figure \ref{fig:Split-Process}. The spacing length $\Delta X_H$ of the new, small particles is half of $\Delta X_L$ of the original, big particles. Hence, the compact support $\epsilon$ of the new, small particles is also half of that of the original, big particles; their volume $V=\Delta x^2$ is $\frac{1}{4}$ of the original, big particles. There are two caveats herein. First, if a boundary has a sharp corner, we always put a particle (black in Figure \ref{fig:Split-Process}) at the corner to ensure accurate representation of the boundary. This corner particle maintained its compact support $\epsilon$ and volume $V$ the same as those of its two neighbor boundary particles. That says, if its neighbor boundary particles were refined, this corner particle synchronized its $\epsilon$ and $V$ with them.  
Second, due to the meshless nature of the GMLS discretization, the interior GMLS particles are allowed not perfectly aligned with the boundaries. After several times of refinement, there could occur near a boundary voids of size larger than the average spacing of surrounding interior particles, which would result in larger local numerical errors. To remedy it, if any interior particle near a boundary needed to be refined, we identified the images of this particle and its neighbor interior particles on the opposite side of the boundary and refined those image (imaginary) particles as well. The resulting new, small particles within the solution domain were kept, which could effectively fill the voids. 

The above four steps~\eqref{def:adap-iter} were repeated iteratively at each time step until the total error satisfied:
\begin{equation}
\label{equ:Error_TOL}
E_{tot} \le TOL \; ,
\end{equation}
where $TOL$ is the preset tolerance.

%% file: results.tex
\section{Simulation results}\label{sec:Simu_results}
Two div-grad problems and four Stokes flow problems were studied to assess the accuracy and convergence of the proposed spatially adaptive GMLS method. 

\subsection{Constant coefficient div-grad problem in an L-shaped domain}\label{subsec:LshapedDomain}

\begin{figure}[htbp]
	\centering
	\includegraphics[width=2.0in]{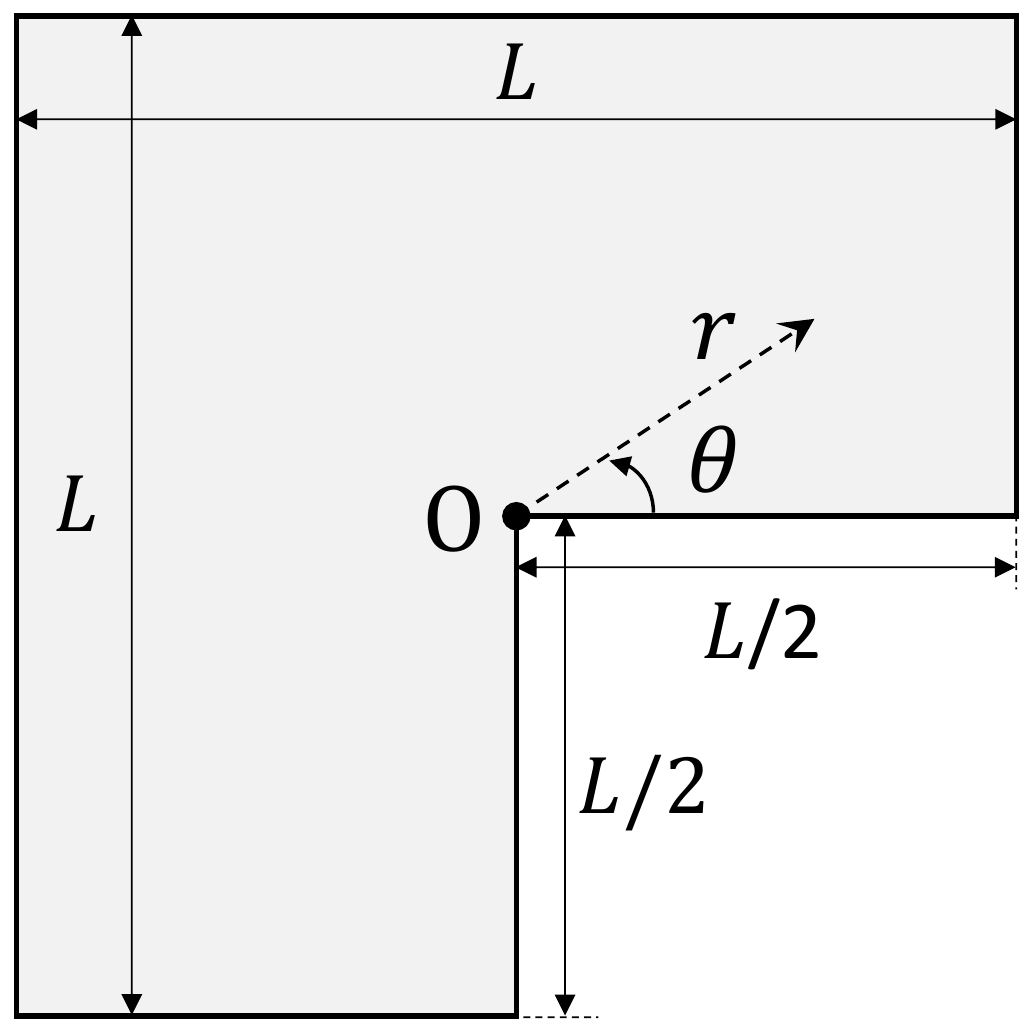}	
	\caption{Geometry of an L-shaped domain with $L=2$. Polar coordinates $(r,\theta)$ are defined at $O$.} 
    \label{fig:LshapedDomain-Geometry}
\end{figure}

We first considered the div-grad problem with $\kappa \equiv 1$ in Eq. \eqref{equ:Elliptic} on an L-shaped domain. This benchmark problem was also studied in a finite element context \cite{schiff1988finite,nochetto2009theory} and isogeometric analysis \cite{johannessen2014isogeometric}. The geometry of the solution domain is described in Figure \ref{fig:LshapedDomain-Geometry}. The source term in Eq. \eqref{equ:Elliptic} was taken as $f=1$. Under polar coordinates, the analytical solution of this problem is \cite{nochetto2009theory}: $ \phi(r,\theta) = r^{\frac{2}{3}}\sin(\frac{2\theta}{3}) - \frac{r^2}{4}$.  

\begin{figure}[htbp]
	\centering
    \subfigure[2nd order polynomial reconstruction.]{
    \includegraphics[width=3.6in]{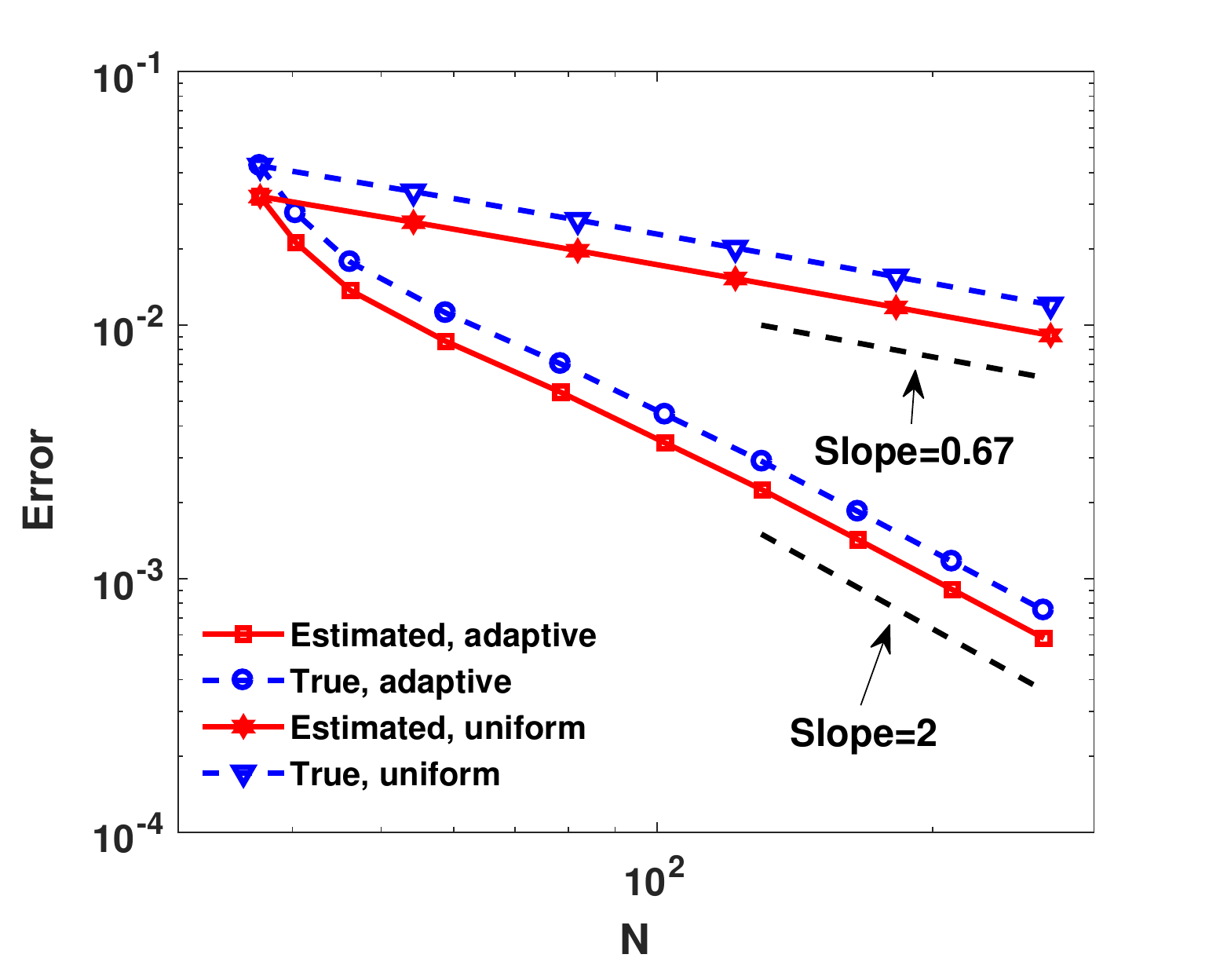}}~
    \subfigure[4th order polynomial reconstruction.]{
    \includegraphics[width=3.6in]{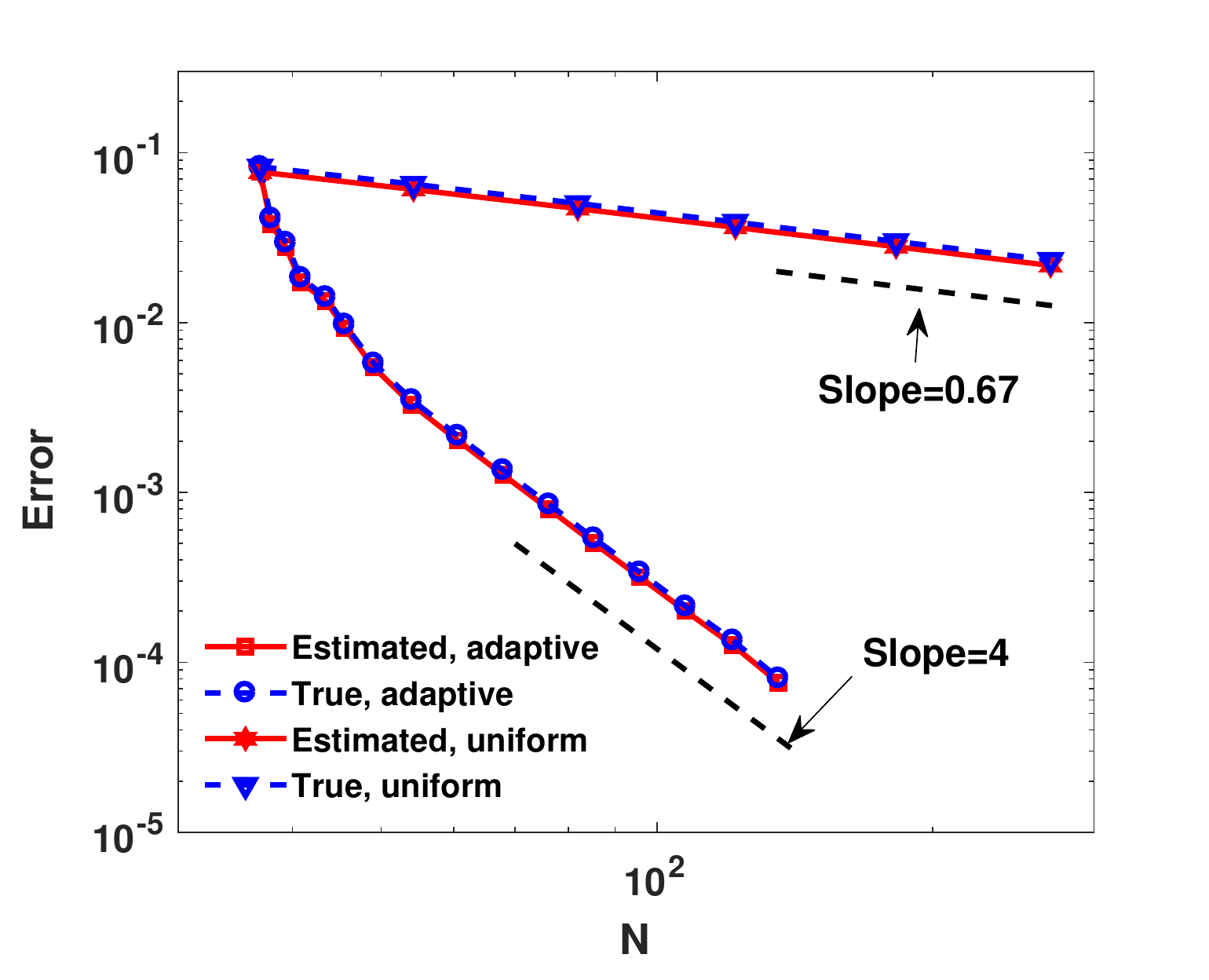}}
	\caption{Convergence of GMLS solution with uniform or adaptive resolution for the div-grad problem on an L-shaped domain. The $x$ axis displays the number of GMLS particles used in each dimension. The $y$ axis is the true or estimated error evaluated by Eq. \eqref{equ:true_error} or \eqref{equ:recover_error}, respectively.} 
    \label{fig:LshapedDomain-Conv}
\end{figure}

Because the L-shaped geometry has a corner singularity, the numerical solution of this problem with \emph{uniform} discretization cannot achieve the theoretical order of convergence, as demonstrated in \cite{nochetto2009theory}. Indeed, we identified the same behavior in our simulations performed with a uniform distribution of GMLS particles. As shown in Figure \ref{fig:LshapedDomain-Conv}, regardless of the polynomial order employed in the GMLS approximation, the convergence order of the numerical solutions remains around 0.67, consistent with reported FEM results \cite{nochetto2009theory}. 

\begin{figure}[htbp]
	\centering
	\includegraphics[width=6in]{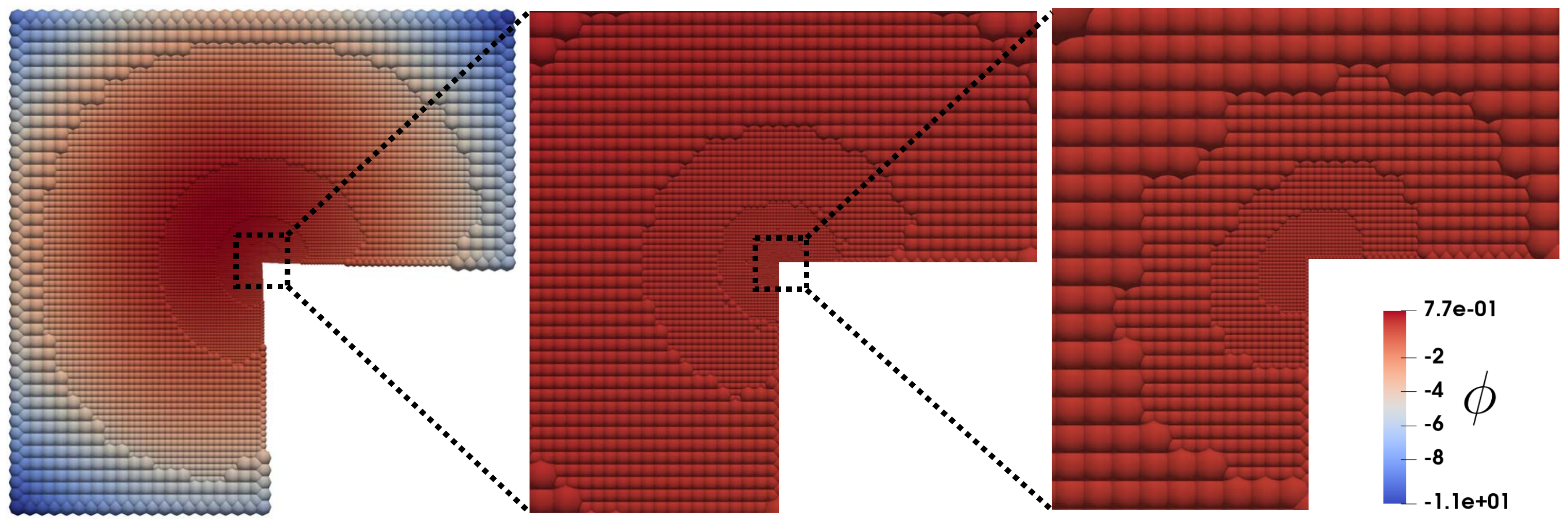}
	\caption{Distribution of GMLS particles with adaptive refinement. Here, the 4th-order polynomial reconstruction was employed. The total number of GMLS particles is 14646. The color is correlated to the value of $\phi$.} 
    \label{fig:LshapedDomain-FinalDis}
\end{figure}

Instead of refining the discretization uniformly, we employed adaptive refinement as proposed in Section \ref{subsubsec:Adapt_Algo}. Figure \ref{fig:LshapedDomain-FinalDis} presents a snapshot of particle distribution under adaptive refinement. As anticipated, the maximum errors occurred near the singular corner of the L-shaped domain. As a result, the GMLS particles around that corner were refined for multiple times to reduce the errors. During these iterations of refinement, the generated adaptive resolution entailed much faster convergence of the numerical solutions, as depicted in Figure \ref{fig:LshapedDomain-Conv}. Notably, e.g., from Figure \ref{fig:LshapedDomain-Conv} (b), to achieve the same accuracy, employing adaptive refinement can significantly reduce the number of GMLS particles needed for solving this problem. 
Furthermore, the adaptive GMLS solution exhibits second or fourth order convergence for second or forth order polynomials used in the GMLS reconstruction, matching the convergence rates for smooth solutions in the literature \cite{Trask2017high}.

In addition, we note that the estimated errors by Eq. \eqref{equ:recover_error} based on the recovered gradient of $\phi$ display the same convergence order as the true errors, in both uniform-resolution and adaptive-resolution scenarios. 
 
\subsection{Div-grad problem with rough coefficients}\label{sec:Elliptic}
We next solved the div-grad problem with a discontinuous coefficient $\kappa$ in Eq. \eqref{equ:Elliptic}, which describes various physical phenomena, e.g., transport in porous media with discontinuous permeability \cite{DGPorousMedia2010,MultiscaleFEMPorousMediaHOU1997,DGPorousMedia1998,PorousMedia1991} and electrostatics with discontinuous dielectric permittivity \cite{Electrostatics2013,luo2010modeling,Electrostatics1991}. The solution domain of this problem is a square of dimension $10 \times 10$. A circle of radius $2$ positioned at the center divides the square into two sub-domains: inside the circle ($\Omega_{f,1}$) and outside the circle ($\Omega_{f,2}$). The coefficient $\kappa$ in Eq. \eqref{equ:Elliptic} is discontinuous across the two sub-domains and given as:
\begin{equation}
\kappa = 
\begin{cases}
30 ~~~& \text{for}~~~ \textbf{x}\in \Omega_{f,1}\\
1 ~~~& \text{for}~~~ \textbf{x}\in \Omega_{f,2} \; 
\end{cases}.
\label{equ:Elliptic_kappa}
\end{equation}
The source term in Eq. \eqref{equ:Elliptic} was taken as $f=0$. To compare with the analytical solution reported in \cite{luo2010modeling} obtained in an infinite domain, the value of $\phi_{\Gamma}$ in Eq. \eqref{equ:Elliptic} was imposed on the boundary of the truncated square domain as a Dirichlet BC \cite{luo2010modeling}. 

\begin{figure}[H]
\centering
\includegraphics[width=4.0in]{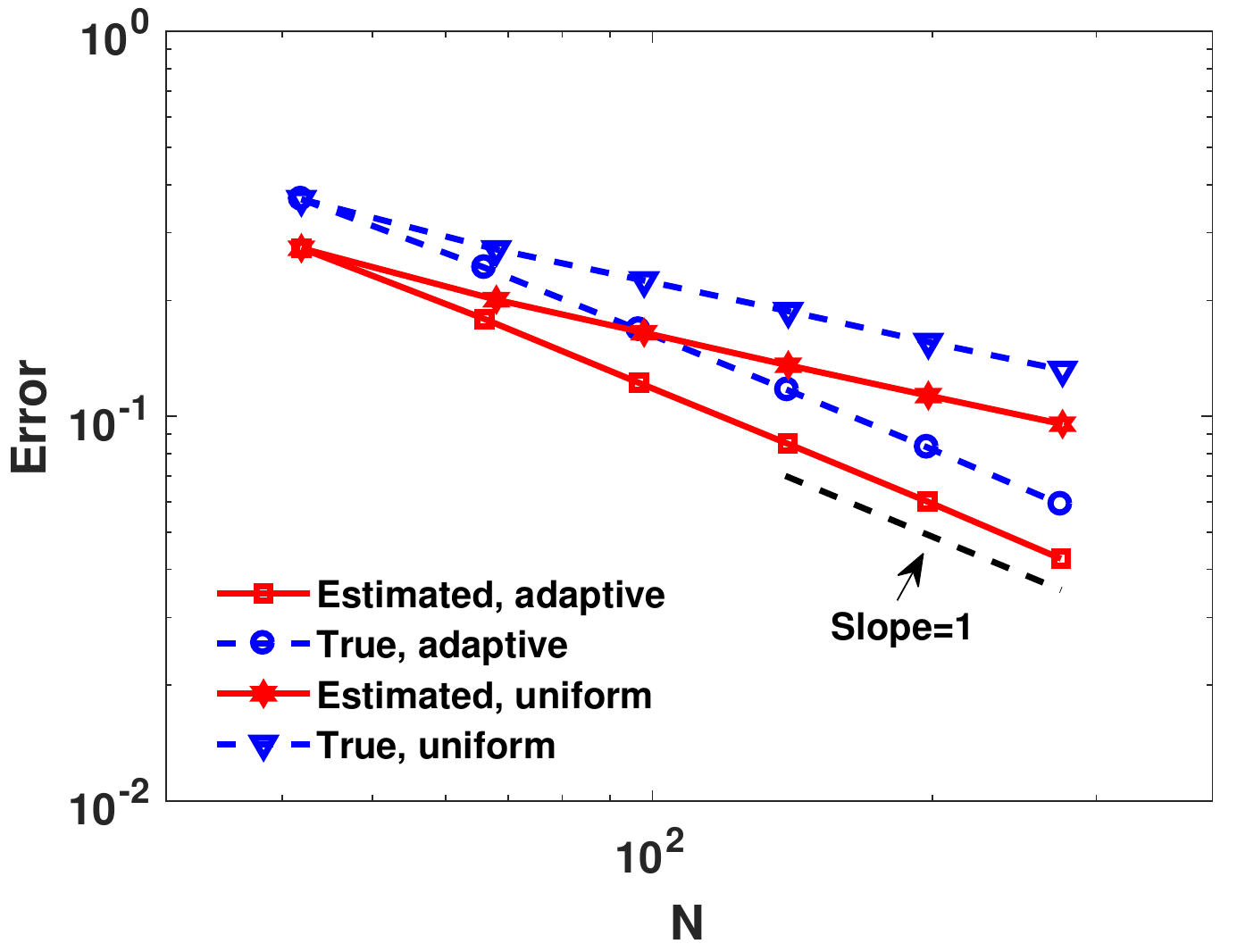}
\caption{Convergence of GMLS solution with uniform or adaptive resolution for the div-grad problem with rough coefficients. The $x$ axis displays the number of GMLS particles used in each dimension. The $y$ axis is the true or estimated error evaluated by Eq. \eqref{equ:true_error} or \eqref{equ:recover_error}, respectively.}
\label{fig:Elliptic-Conv}
\end{figure}

Provided such a large coefficient ratio between the two adjacent sub-domains as in Eq. \eqref{equ:Elliptic_kappa}, the numerical solution near the interface tends to develop oscillations, and hence, it's challenging to solve this problem stably and accurately. Trask et al. has previously demonstrated that the compatible, staggered GMLS discretization can effectively eliminate numerical oscillations and produce stable and accurate solutions for this problem \cite{Trask2017high}. However, we find that due to the discontinuous coefficient, the convergence of the numerical solution is diminished when using uniformly distributed GMLS particles. As depicted in Figure \ref{fig:Elliptic-Conv}, the true errors of the numerical solutions with uniform refinement exhibit less than 1st-order convergence. Here, to avoid the Runge phenomenon \cite{epperson1987runge,michael2002scientific}, the 1st-order polynomial basis was employed in the reconstruction of the gradient, which theoretically entails 1st-order convergence. 

\begin{figure}[htbp]
\centering
\includegraphics[width=7.0in]{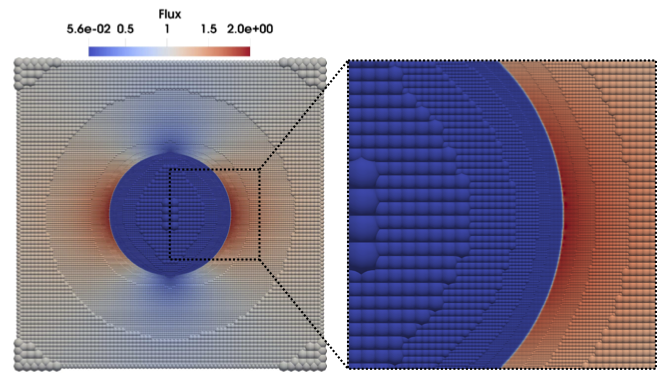}
\caption{Distribution of GMLS particles with adaptive refinement. The color is correlated to the magnitude of $\partial\phi/\partial x$.}
\label{fig:Elliptic-Distribution}
\end{figure}

\begin{figure}[htbp]
\centering
\includegraphics[width=4.0in]{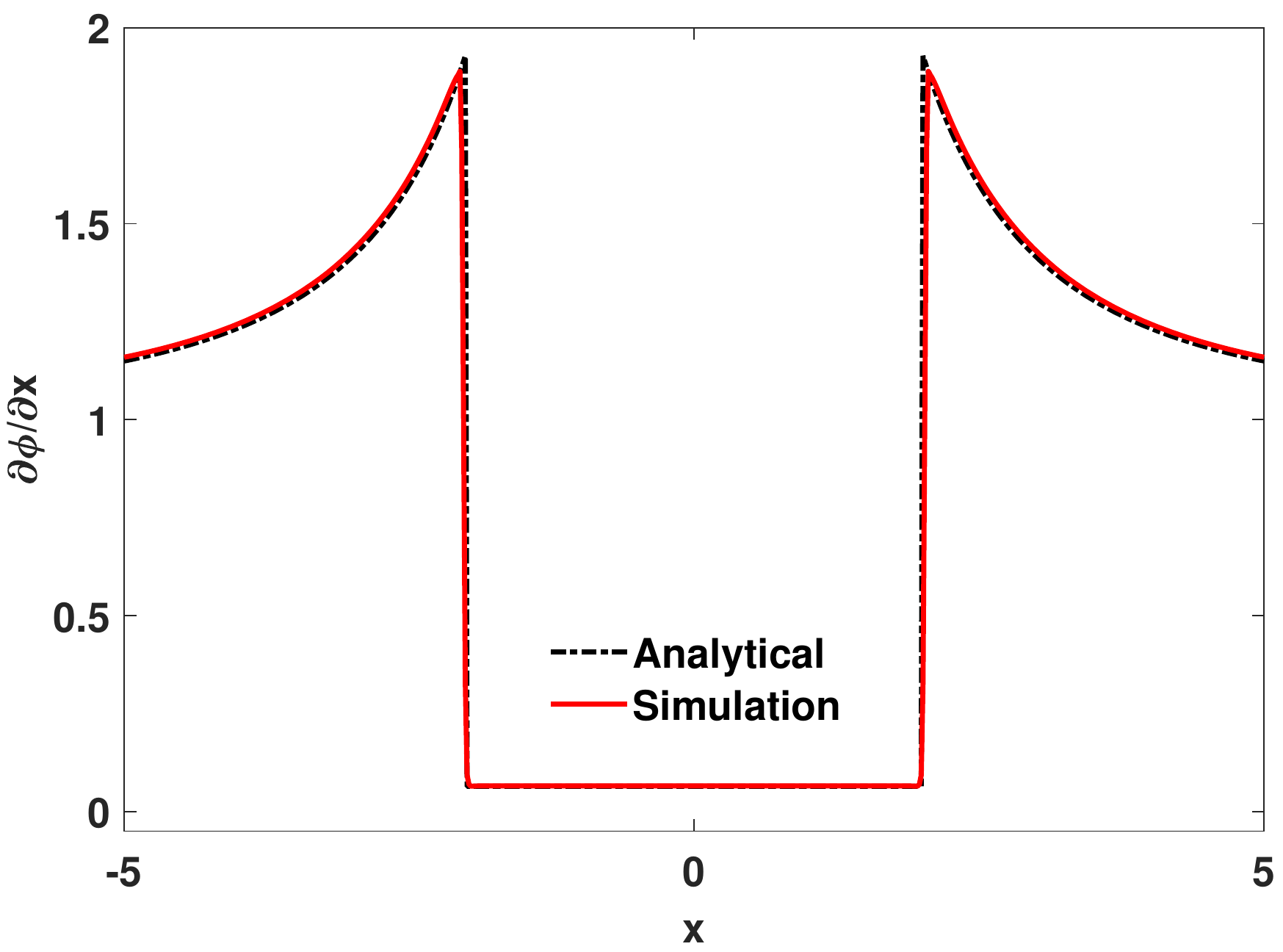}
\caption{$\frac{\partial \phi}{\partial x}$ along the line $y=0$ computed by adaptive GMLS and compared with analytical solution \cite{luo2010modeling}.}
\label{fig:Elliptic-Flux}
\end{figure}

Thus, we demonstrate herein that the optimal convergence order can be achieved by introducing adaptive refinement in the staggered GMLS discretization. Figure \ref{fig:Elliptic-Distribution} illustrates the distribution of GMLS particles after several iterations of refinement. The proposed error estimator directed the adaptive refinement with the finest resolution near the interface of the two sub-domains.
Figure \ref{fig:Elliptic-Flux} presents the computed $\frac{\partial \phi}{\partial x}$ (flux along $x$) along the line ($y=0$) passing through the center of the square domain. The numerical solution agrees well with the analytical solution \cite{luo2010modeling}. The convergence study shown in Figure \ref{fig:Elliptic-Conv} confirms that the \emph{adaptive} GMLS solution can achieve the theoretical 1st-order convergence. 
The estimated errors based on the recovered gradient of $\phi$ display the same convergence order as the true errors, in both uniform-resolution and adaptive-resolution scenarios.

\subsection{Wannier flow}\label{subsec:wannier}

We now move forward to the Stokes problems. In this section, we examined Wannier flow \cite{wannier1950contribution}, a Stokes flow confined between two rotating cylinders. Wannier flow has its analytic solutions for both pressure and velocity, providing a good benchmark for assessing accuracy and convergence of the proposed method. GMLS was previously employed for solving Wannier flow with the convergence assessed for uniform resolution \cite{Trask2018compatible}.  

\begin{figure}[htbp]
	\centering
	\includegraphics[width=3.0in]{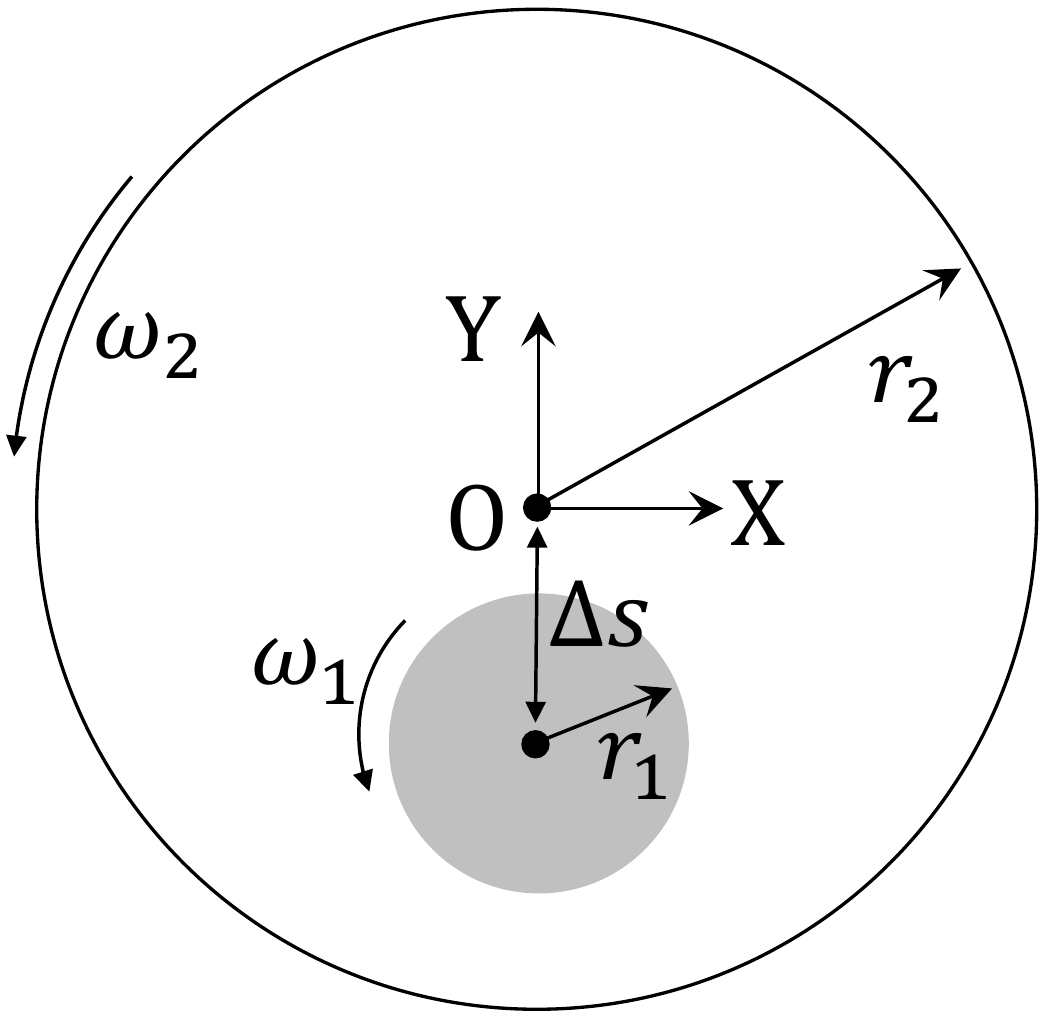}	
	\caption{Setup of Wannier flow problem with $r_1=\frac{\pi}{10}~\si{m}$, $r_2=\frac{\pi}{2}~\si{m}$, $\omega_1 = \frac{10}{\pi}~\si{s}^{-1}$, and $\omega_2 = \frac{1}{\pi}~\si{s}^{-1}$.} 
    \label{fig:Wannier-Geometry}
\end{figure}

The specific setup of the problem is sketched in Figure \ref{fig:Wannier-Geometry}, where the radii of the two cylinders are $r_1$ and $r_2$, respectively; the distance between the two cylinders' centers is defined as $\Delta S$; and the two cylinders are rotating in the same direction with the rotational speeds $\omega_1$ and $\omega_2$, respectively. The fluid was assumed to have the density $\rho = 1.0~\si{kg/m^3}$ and kinematic viscosity $\nu = 1.0~\si{m^2/s}$.

\begin{figure}[!ht]
\centering
\subfigure[2nd order polynomial reconstruction.]{
\includegraphics[width=3.6in]{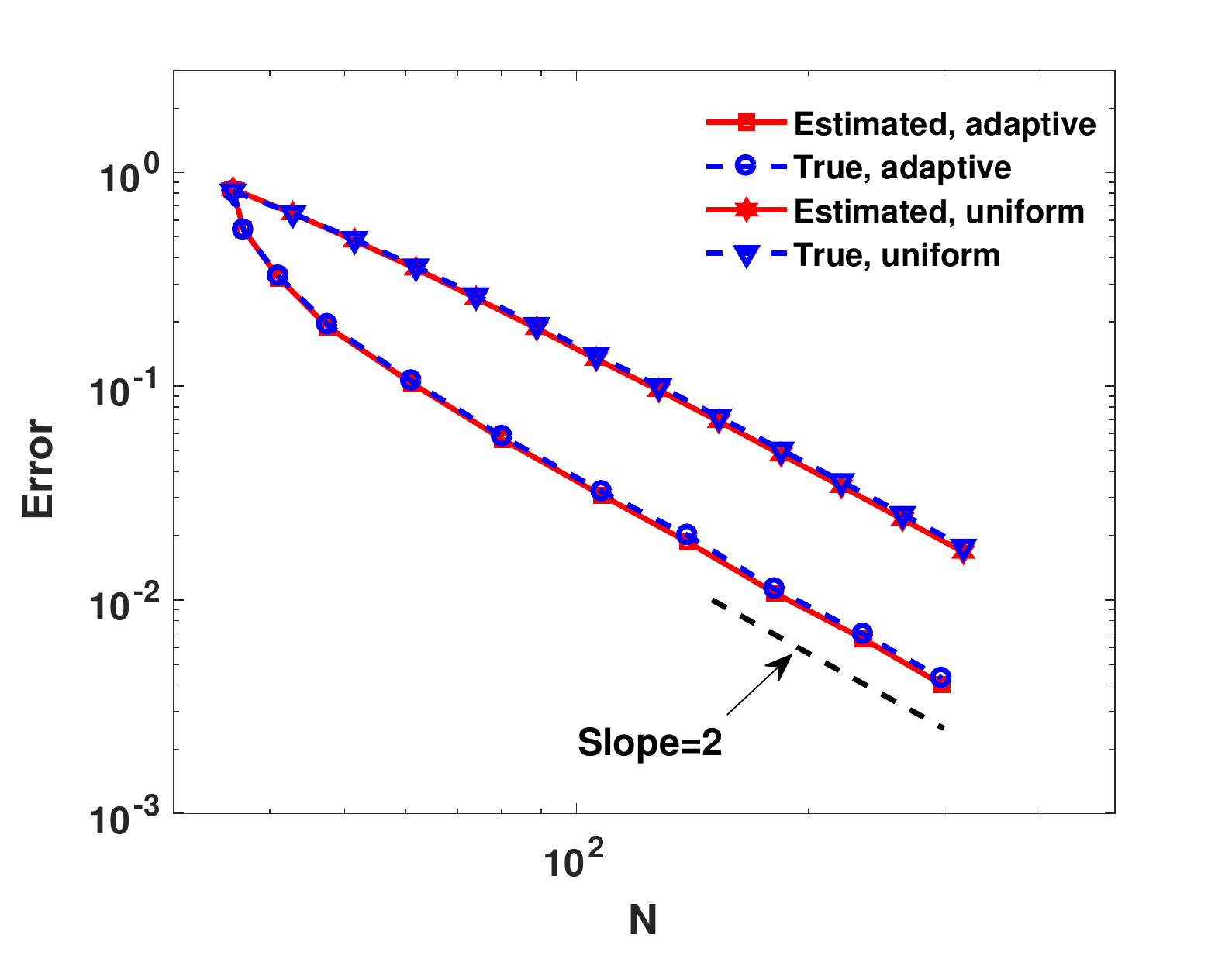}}~
\subfigure[4th order polynomial reconstruction.]{
\includegraphics[width=3.6in]{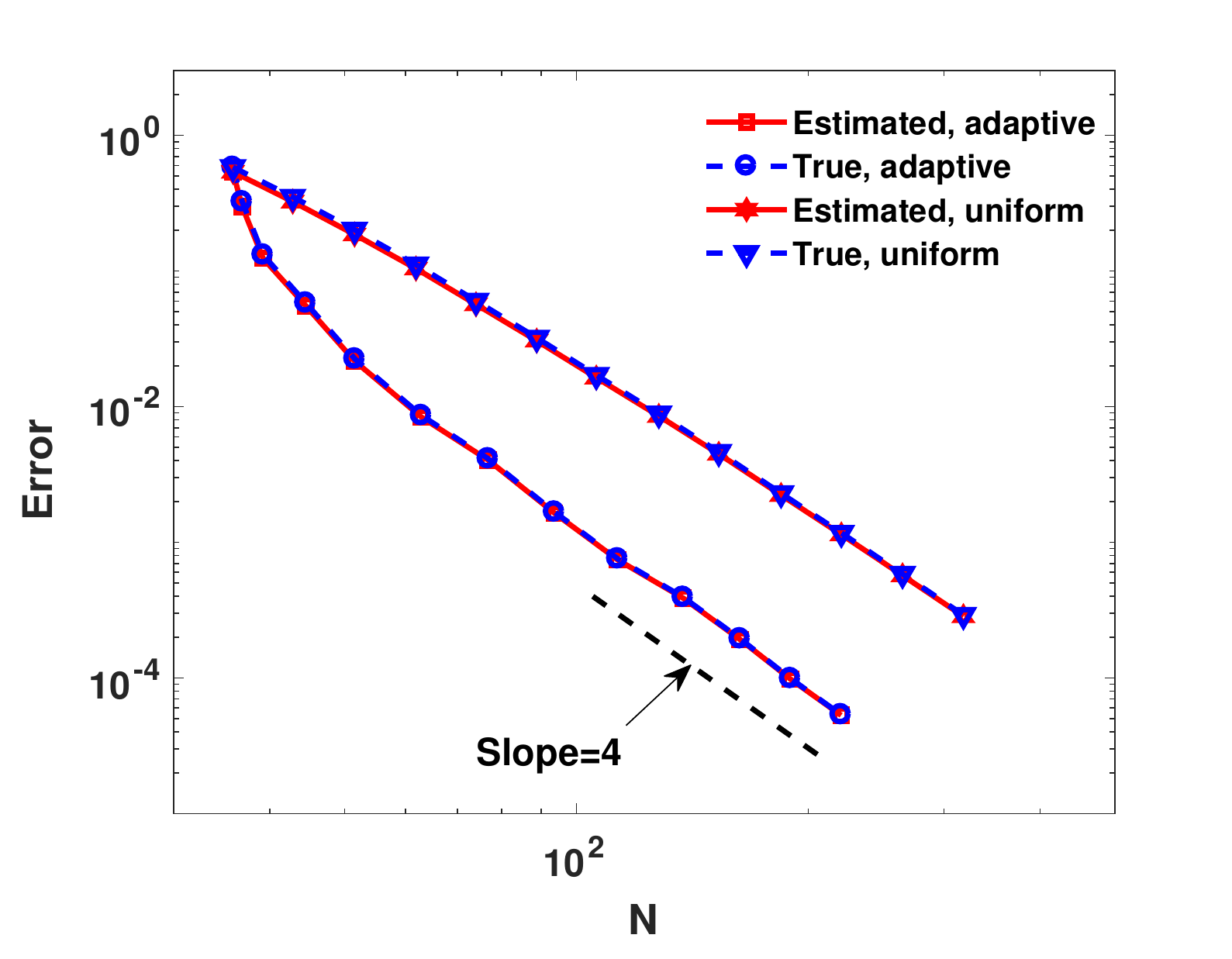}}
\caption{Convergence of GMLS solution for Wannier flow with adaptive or uniform refinement. The $y$ axis is the true or estimated recovery-based error for the velocity.}
\label{fig:Wannier-Conv}
\end{figure}

Starting with a coarse, uniform distribution of GMLS particles, we again followed the algorithm described in Section \ref{sec:adaptive} to apply adaptive refinement. During the process, the marked GMLS particles were split, and hence, the total number of GMLS particles ($N^2$) increased. In each iteration of refinement, the true and estimated recovery-based errors were evaluated, as depicted in Figure \ref{fig:Wannier-Conv}. It can be seen that the optimal convergence order is obtained for the gradient of velocity, regardless of 2nd or 4th-order polynomial reconstruction employed in GMLS. For comparison, we also employed uniform refinement and evaluated the true and estimated errors, also depicted in Figure \ref{fig:Wannier-Conv}. Notably, using the same number of GMLS particles, the adaptive refinement yield more accurate solutions than uniform refinement. Given the same acceptable error, employing adaptive refinement would require many fewer degrees of freedom in the simulation. Also, we note that the estimated errors by Eq. \eqref{equ:recover_error} based on the recovered gradient of velocity closely match the true errors, in both uniform and adaptive refinement. In this specific test, we set $\Delta S=\frac{\pi}{5}~\si{m}$. 
%

\begin{figure}[!ht]
	\centering
	\includegraphics[width=6in]{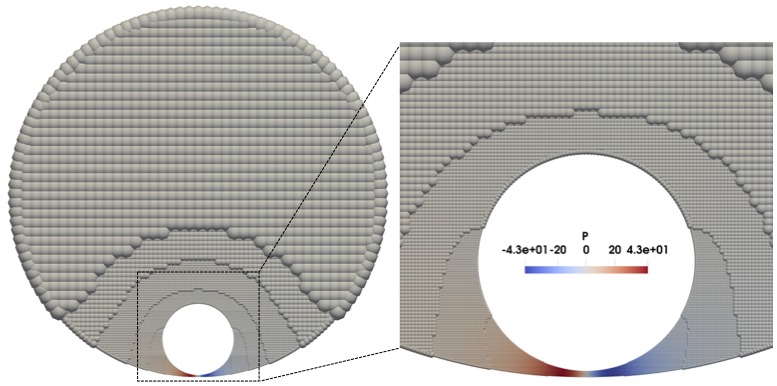}	
	\caption{Distribution of GMLS particles with adaptive refinement when the minimum gap width between the two cylinders $r_2-r_1-\Delta S = \frac{r_1}{20}$. The number of GMLS particles is 18138. The color is correlated to the computed pressure (Unit in color bar: Pa).} 
    \label{fig:Wannier-Pressure}
\end{figure}

\begin{figure}[!ht]
	\centering
	\includegraphics[width=4in]{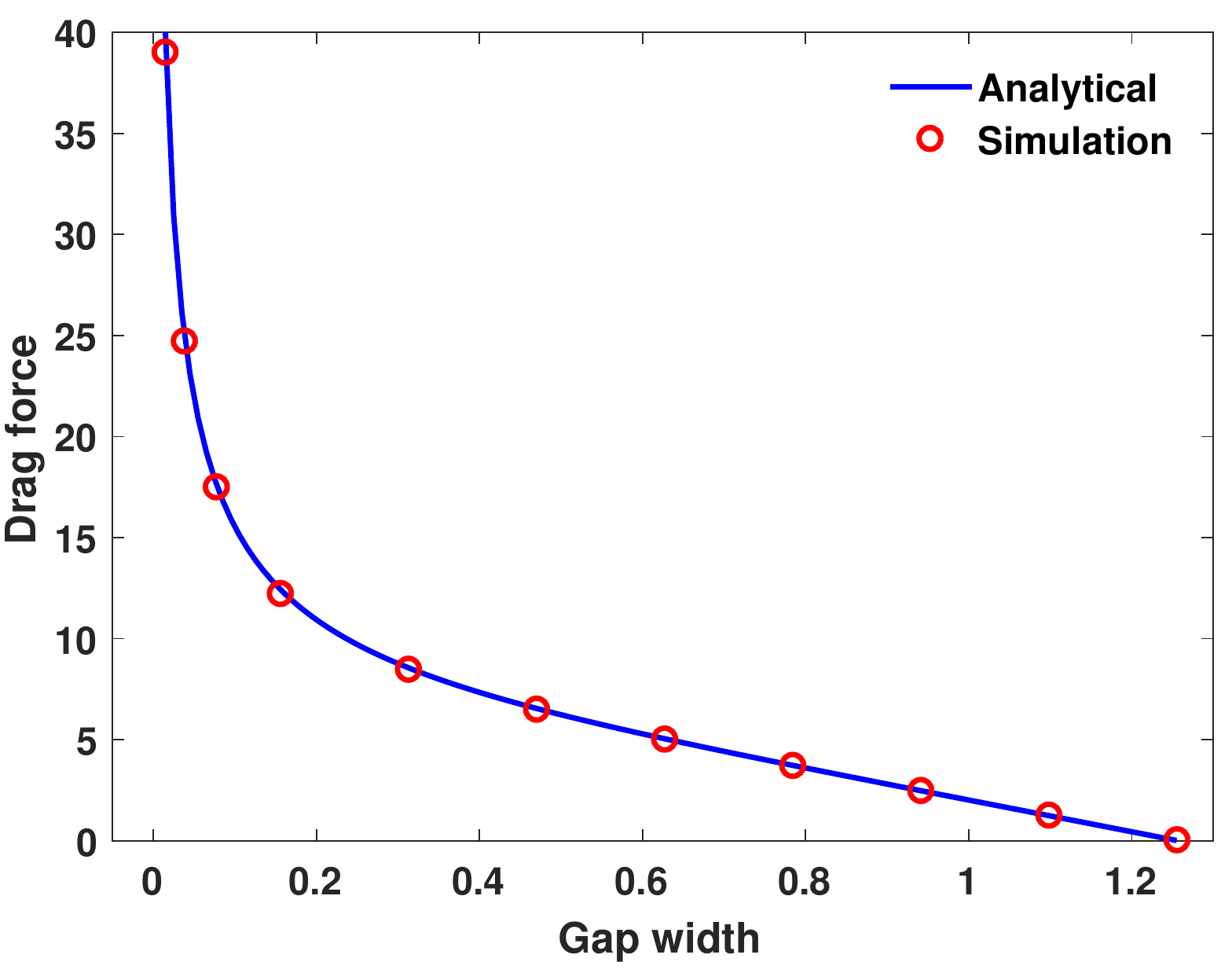}	
	\caption{Drag force (Unit: $\si{N}$) exerted on the inner cylinder under Wannier flow as a function of gap width (Unit: $\si{m}$) between the two cylinders, computed by adaptive GMLS and compared with analytical solution \cite{wannier1950contribution}.} 
    \label{fig:Wannier-Drag}
\end{figure}

By adjusting the center position of the inner cylinder, the minimum gap width between the two cylinders ($r_2-r_1-\Delta S$) varies. As the two cylinders draw close, a numerical solution could fail due to lack of sufficient discretization points within the narrow gap between the two cylinders. In this case, adaptive refinement is necessary to obtain an accurate and computationally tractable numerical solution. Figure \ref{fig:Wannier-Pressure} depicts the resulting distribution of GMLS particles after 6 iterations of refinement. Here, the minimum gap width between the two cylinders ($r_2-r_1-\Delta S$) is as small as $\frac{r_1}{20}$. From Figure \ref{fig:Wannier-Pressure}, we see that adaptation allows for sufficient discretization points within the narrow gap, thereby enabling a stable and accurate numerical solution in such an extreme case. The computed pressure distribution shown in Figure \ref{fig:Wannier-Pressure} agrees with the analytical solution \cite{wannier1950contribution}. 
Further, we computed the drag force exerted on the inner cylinder by the fluid at varying gap widths. As shown in Figure \ref{fig:Wannier-Drag}, the numerical predictions show good agreement with the analytical solution \cite{wannier1950contribution} for all cases considered. For the results presented in both Figure \ref{fig:Wannier-Pressure} and Figure \ref{fig:Wannier-Drag}, the 4th-order GMLS polynomial reconstruction was employed for both the velocity and pressure. 

\subsection{Dynamics of an L-shaped active colloid}\label{subsec:LshapedColloid} 

\begin{figure}[htbp]
	\centering
	\includegraphics[width=5in]{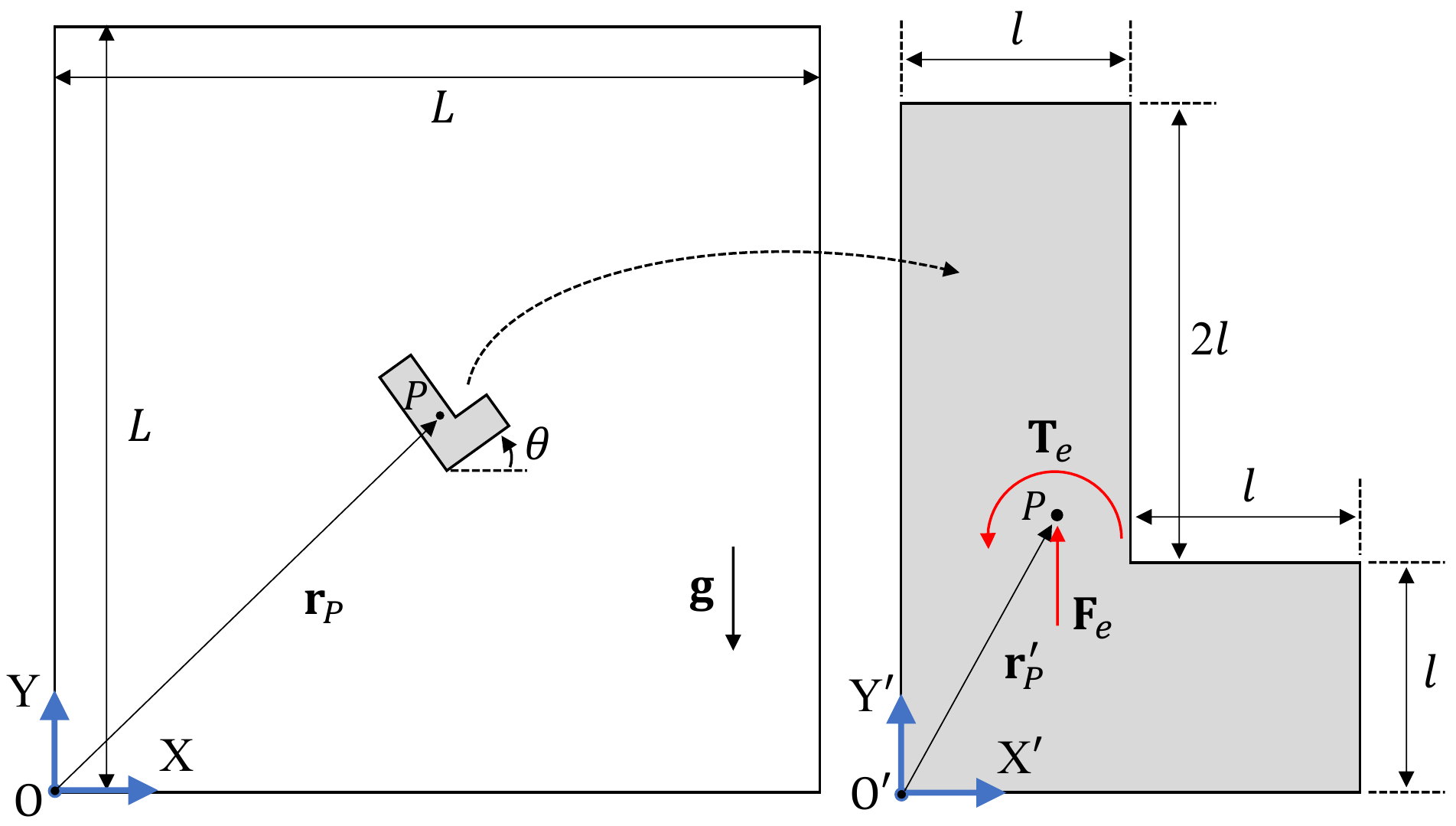}	
	\caption{Configuration of the solution domain and L-shaped colloid with $L = 40\times10^{-6}~\si{m}$ and $l = 2\times10^{-6}~\si{m}$.} \label{fig:Lshape-Geometry}
\end{figure}

In this section, we reproduced an interesting phenomenon concerning the motion of an asymmetric self-propelled colloid in fluid, which was first observed experimentally \cite{ten2014gravitaxis}. The colloid is L-shaped with two asymmetric arms, and its motion is constrained in 2D, as in the experiment \cite{ten2014gravitaxis}. In our simulation, the L-shaped colloid was suspended in a square domain chosen large enough that the boundaries had insignificant effects on the colloid dynamics. As the colloid moved, we always ensured that the colloid's center of mass (COM) coincided with the center of the square domain, as depicted in Figure \ref{fig:Lshape-Geometry}. 
The BCs specified in Eq. \eqref{equ:StokesEqn2} were enforced on the boundaries of the square domain and L-shaped colloid. The colloid was subject to the gravitational force $\mathbf{g}$ {with $\|\mathbf{g}\|= 0.5\times10^{-9}~\si{N}$}. To mimic the colloid's self-propelling mechanism as in the experiment, the equivalent force $\textbf{F}_e$ and torque $\textbf{T}_e$ were applied at the colloid's COM. During the motion of the colloid, $\textbf{F}_e$ was exerted always along the longer arm of the colloid. 
The torque $\textbf{T}_e$ was related to $\textbf{F}_e$ by $\textbf{T}_e = (l - x'_p) \textbf{e}_{\textbf{x}'} \times \textbf{F}_e$, where $x'_p$ and $\textbf{e}_{\textbf{x}'}$ were defined in the local coordinate system on the colloid with the origin $O'$ set at the bottom left corner and $\textbf{e}_{\textbf{x}'}$ along the shorter arm. The coordinates of the colloid's COM $(x'_p,~y'_p) = (1.98\times10^{-6}~\si{m},~2.02\times10^{-6}~\si{m})$. The fluid assumed the properties of water: $\rho = 1.0\times 10^3~\si{kg/m^3}$ and $\nu = 1.0\times 10^{-6}~\si{m^2/s}$, consistent with the experiment \cite{ten2014gravitaxis}.

As found in the experiment \cite{ten2014gravitaxis}, by varying the magnitude of $\textbf{F}_e$ (and thereby the magnitude of $\textbf{T}_e$ as well), the L-shaped active colloid can display rich dynamic behaviors owing to its nonlinear hydrodynamic interaction with the surrounding fluid. As a result, the colloid can follow different types of trajectories with different magnitudes of $\textbf{F}_e$. More specifically, for a small $\textbf{F}_e$, since the colloid's self-propelling motion is slow, and the impact of gravitational force is dominant, the colloid exhibits straight downward swimming (SDS). As $\textbf{F}_e$ gradually increases, the colloid's active motion becomes more pronounced, and its hydrodynamic interaction with the surrounding fluid starts to compete with the gravitational force, leading to straight upward swimming (SUS) or even trochoid-like motion (TLM). Because the associated fluid dynamics falls into the low Reynolds number regime \cite{ten2014gravitaxis}, it may be treated as a Stokes problem.

\begin{figure}[htbp]
	\centering
	\includegraphics[width=4in]{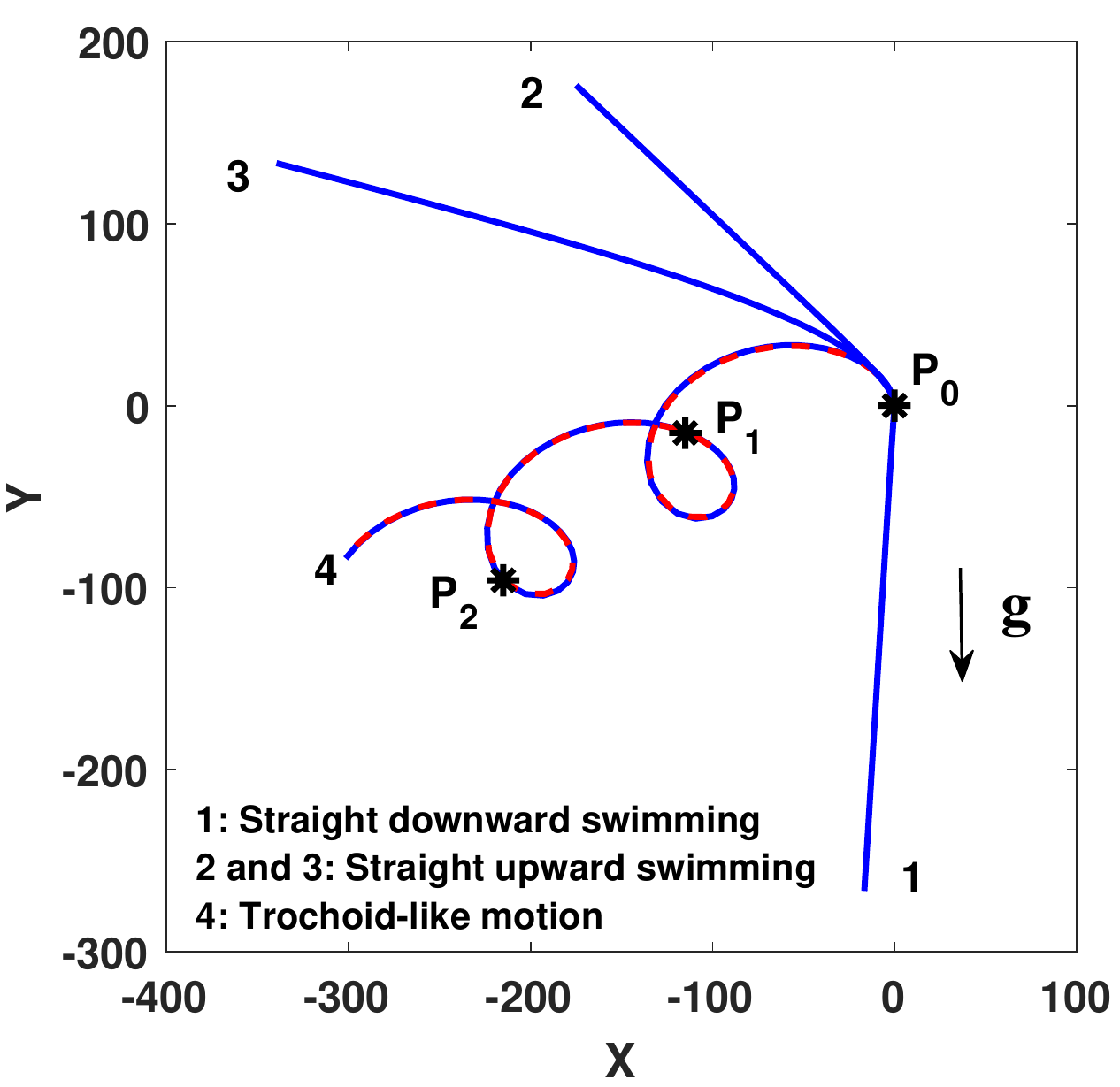}
	\caption{Typical trajectories (Unit: $10^{-6}~\si{m}$) of the L-shaped active colloid with different magnitudes of $\textbf{F}_e$. Specifically, $||\textbf{F}_e||=0.0$ for Trajectory 1; $||\textbf{F}_e||=1.5,~2.0\times10^{-9}~\si{N}$ for Trajectory 2 and 3, respectively; and, Trajectory 4 corresponds to $||\textbf{F}_e||=3.0\times10^{-9}~\si{N}$. All of the trajectories started from the same initial position ($P_0$). For Trajectory 4, the prediction by adaptive GMLS (blue solid line) is compared with the result of GMLS with the uniform, high resolution (red dash line).}
	\label{fig:Lshape-Trajectory}
\end{figure}

\begin{figure}[htbp]
\centering
\subfigure[Snapshot at the time corresponding to $P_1$ on the TLM trajectory in Figure \ref{fig:Lshape-Trajectory}.]{
\includegraphics[width=6.0in]{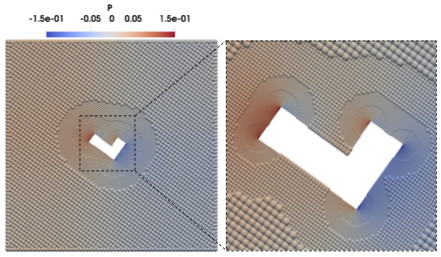}}
\subfigure[Snapshot at the time corresponding to $P_2$ on the TLM trajectory in Figure \ref{fig:Lshape-Trajectory}.]{
\includegraphics[width=6.0in]{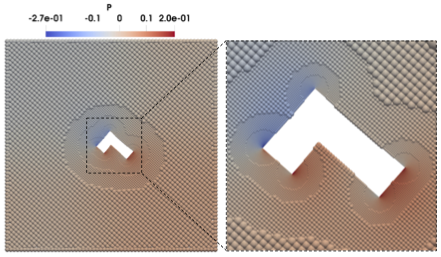}}
\caption{Two snapshots showing the distributions of GMLS particles with adaptive refinement at the two different times corresponding to $P_1$ and $P_2$, respectively, marked on the TLM trajectory shown in Figure \ref{fig:Lshape-Trajectory}. The color is correlated to the computed fluid pressure (Unit in color bar: $10^{-9}~\si{Pa}$).}
\label{fig:Lshape-pressure}
\end{figure}

To solve the Stokes equations  \eqref{equ:StokesEqn2} and \eqref{equ:ForceTorque_Free}, we employed the proposed adaptive GMLS method. The numerical solutions captured all three typical types of motions, as illustrated in Figure \ref{fig:Lshape-Trajectory}. In Figure \ref{fig:Lshape-pressure}, we further demonstrate two snapshots of the adaptive GMLS simulation, where the distribution of GMLS particles with adaptive refinement can be seen. The refinement directed by the error estimator concentrated around the corners of the L-shape, which should be anticipated. The two snapshots correspond to the two points ($P_1$ and $P_2$) marked on the TLM trajectory shown in Figure \ref{fig:Lshape-Trajectory}, respectively. (Refer to the animation in the Supporting Information for the entire dynamics of TLM.) In Figure \ref{fig:Lshape-pressure}, we also present the computed pressure field surrounding the colloid. To verify the adaptive GMLS solution, we compared the computed trajectory of the colloid with that predicted by GMLS with a uniform, high resolution. In particular, the most complicated TLM was chosen for verification. As depicted in Figure \ref{fig:Lshape-Trajectory}, the two solutions overlap with each other. Notably in this test, while an average of 13800 GMLS particles were used in the adaptive-resolution GMLS, 58810 GMLS particles were needed in the simulation with the uniform, high resolution. Obviously, the adaptive GMLS required much fewer degrees of freedom for the same accuracy. 
%

\subsection{Dynamics of two square colloids under a shear flow}\label{sec:TwoSquare}

\begin{figure}[htbp]
	\centering
	\includegraphics[width=3in]{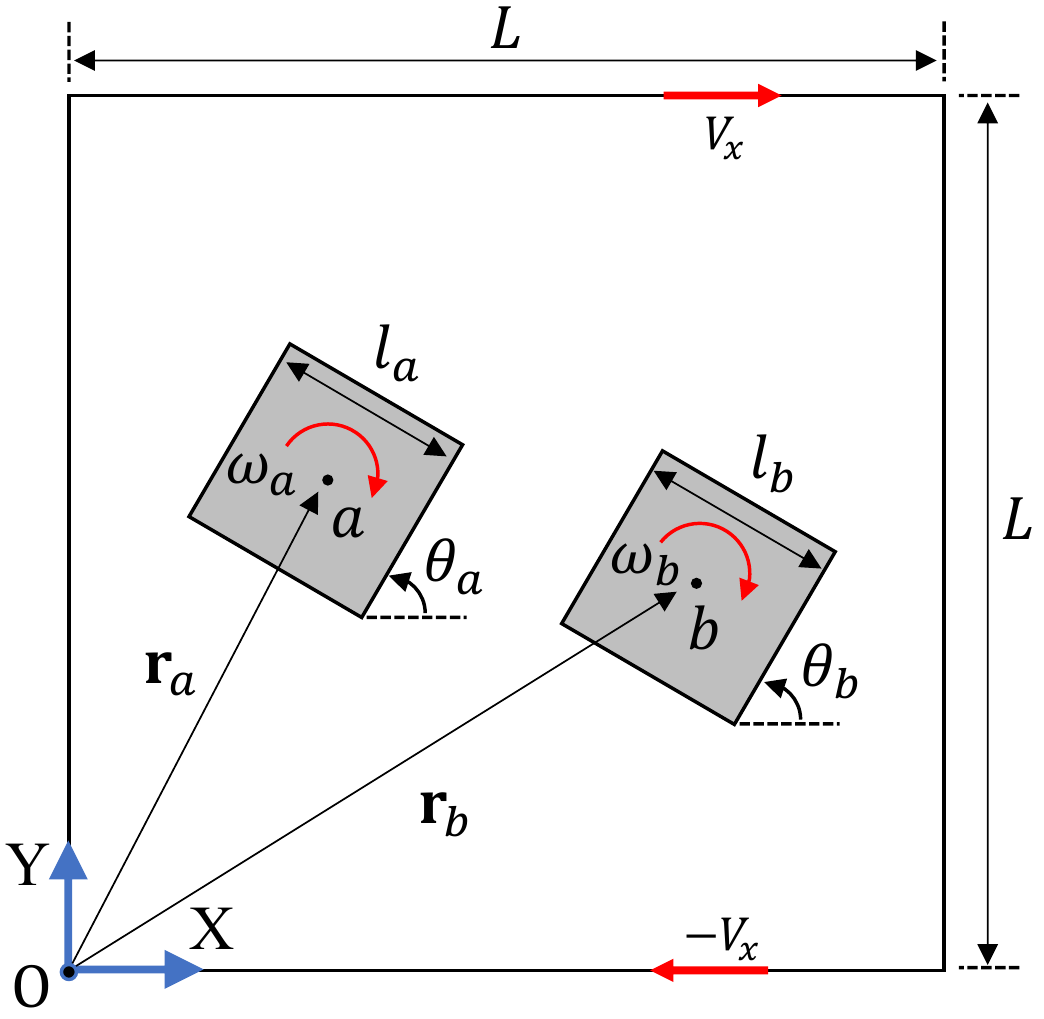}
	\caption{Setup of the Stokes problem with two square colloids under a shear flow. Here, $l_a=l_b=2\times10^{-6}~\si{m}$; $L=40\times10^{-6}~\si{m}$; $V_x=L\dot{\gamma}/2$ with the shear rate $\dot{\gamma}=0.1~\si{s}^{-1}$. The initial positions of the two colloids' COMs relative to the center of the domain are: $(-1\times10^{-6}~\si{m},~0.5\times10^{-6}~\si{m})$ and $(1\times10^{-6}~\si{m},~ -0.5\times10^{-6}~\si{m})$, and the initial orientations of the two colloids are given by $\theta_a = \theta_b = \frac{\pi}{6}$.}
    \label{fig:TwoSquare-Geometry}
\end{figure}

In this section, we solved the Stokes problem with two colloids immersed in fluid. In particular, the two colloids were square-shaped and subject to a shear flow. Figure \ref{fig:TwoSquare-Geometry} illustrates the setup of this problem. Two colloids of equal size were suspended in a square domain large enough to neglect wall effects. The fluid had the same density and kinematic viscosity as in Section \ref{subsec:wannier}. To generate a plane shear flow, the top and bottom walls were assigned velocities consistent with the target shear rate. Following the shear flow, the two colloids rotate while approaching each other and then moving apart. When the two colloids get very close, it can be challenging for a numerical solver to be stable and still provides accurate solutions without invoking a subgrid-scale lubrication model \cite{yuan1994rheology,kromkamp2006lattice,YEOJCP2010}. 

In literature, a similar problem but with two circular colloids has been solved both analytically \cite{darabaner1967particle} and numerically \cite{bian2014splitting,Trask2018compatible,Pan_ARSPH_2019}. Solving the problem with square colloids is more challenging due to the presence of sharp corners causing pressure singularities. To this end, the GMLS discretization enables a sharp-representation of arbitrary shapes, and the adaptive GMLS allows for resolving the singularity with sufficient resolution to ensure stable and accurate solutions. To validate, we also solved the problem with two circular colloids and compare with its analytical solution \cite{darabaner1967particle}. The 4th-order GMLS polynomial reconstruction was employed for both the velocity and pressure.
 
\begin{figure}[!htp]
	\centering
	\includegraphics[width=4.0in]{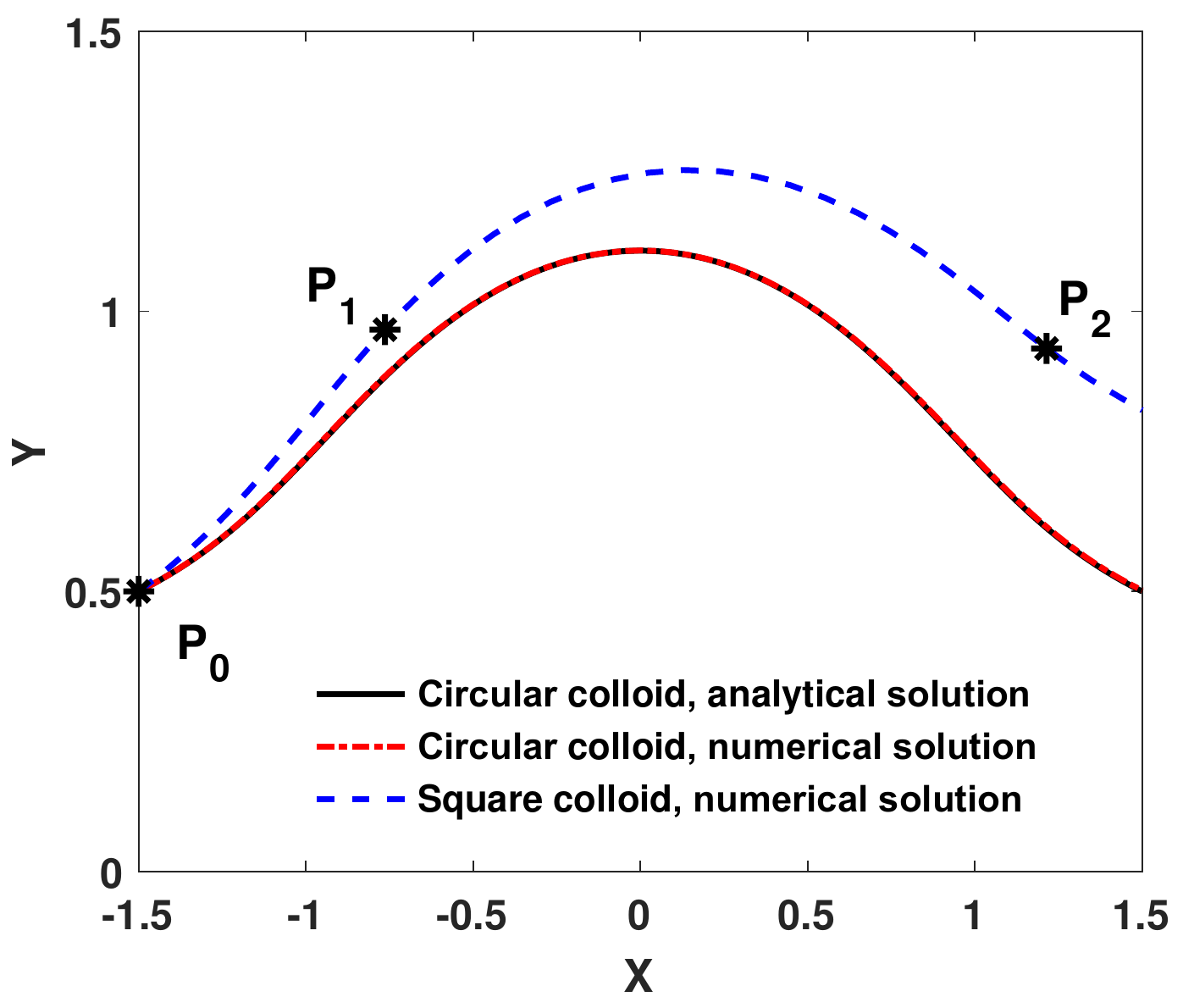}	
	\caption{Trajectory (Unit: $10^{-6}~\si{m}$) of the COM of the colloid initially on the left, computed by adaptive GMLS. For the case of circular colloids, the numerical solution is compared with analytical solution \cite{darabaner1967particle}.} 
    \label{fig:TwoSquare-Trajectory}
\end{figure}

Figure \ref{fig:TwoSquare-Trajectory} depicts the trajectories computed by adaptive GMLS. Due to the symmetry of the two colloids, we only tracked the trajectory of the COM of the colloid initially on the left. As shown in Figure \ref{fig:TwoSquare-Trajectory}, our numerical prediction well agrees with the analytical solution \cite{darabaner1967particle} for the case of circular colloids, which validated the adaptive GMLS solution. We also note that the shape of colloids plays a significant role on their dynamics and trajectories. While the trajectory for the circular colloids was symmetric, the square colloids followed an asymmetric path and did not return to their initial positions. For convenience of comparison, the diameter of the circular colloids was set equal to the square colloids' edge length. 

\begin{figure}[!ht]
\centering
\subfigure[$t=0~\si{s}$ ]{
\includegraphics[width=5.0in]{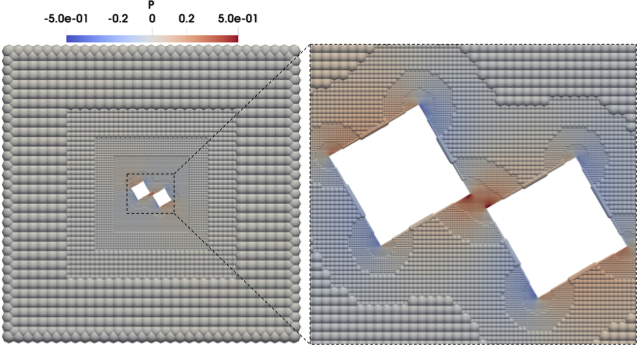}}
\subfigure[$t=20~\si{s}$ ]{
\includegraphics[width=5.0in]{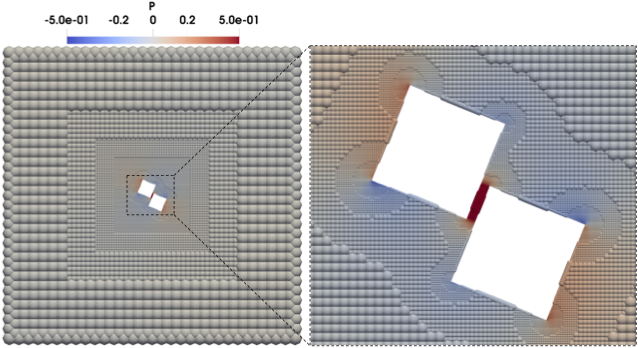}}
\subfigure[$t=45~\si{s}$ ]{
\includegraphics[width=5.0in]{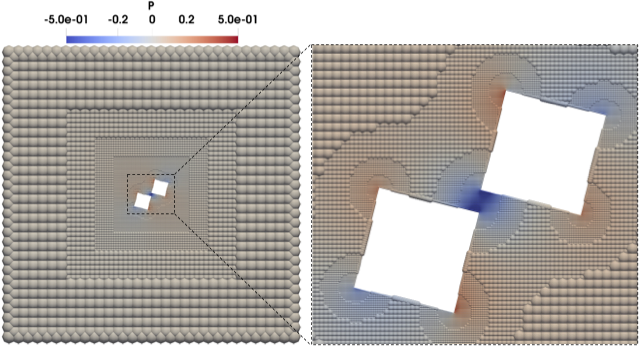}}
\caption{Three snapshots showing the distributions of GMLS particles with adaptive refinement at the three different times corresponding to $P_0$, $P_1$, and $P_2$, respectively, marked on the trajectory shown in Figure \ref{fig:TwoSquare-Trajectory}. Here, an average of 13005 GMLS particles were used. The color is correlated to the computed fluid pressure (Unit in color bar: $10^{-9}~\si{Pa}$).}
\label{fig:TwoSquare-Pressure}
\end{figure}

In Figure \ref{fig:TwoSquare-Pressure}, we present the distributions of GMLS particles at three different times corresponding to $P_0$, $P_1$, and $P_2$, respectively, marked on the trajectory shown in Figure \ref{fig:TwoSquare-Trajectory}. (Refer to the animation in the Supporting Information for more snapshots.) Since the simulation domain is large compared with the size of the colloids, if we started from a uniform, coarse distribution of particles throughout the entire domain, the computation would be still too expensive even with adaptive refinement. Thus, we instead started from an initial multi-resolution scenario. In particular, the simulation domain was divided into four sub-domains discretized with different resolutions depending on their distances to the center of the domain. The furthest subdomain was discretized with the coarsest resolution, and for each two adjacent sub-domains, the resolution ratio was set to 2. With this multi-resolution scenario as the initial distribution of GMLS particles, we applied the adaptive refinement. The error estimator directed refinement near the sharp corners of the square colloids (see Figure \ref{fig:TwoSquare-Pressure}). The computed pressure field is also presented in Figure \ref{fig:TwoSquare-Pressure}, from which we can see that the 4th-order GMLS solution with adaptive refinement is able to stably and correctly capture the short-range lubrication interaction between the two square colloids even when they get very close. Here, no subgrid-scale lubrication model or correction was used.

\subsection{Collective dynamics of active asymmetric colloidal dimers}\label{subsec:AsymmetricColloids} 

\begin{figure}[!htp]
\centering
\subfigure[]{
\includegraphics[width=5.0in]{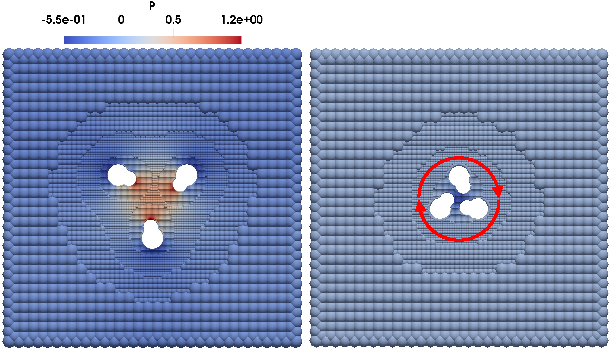}}
\subfigure[]{
\includegraphics[width=5.0in]{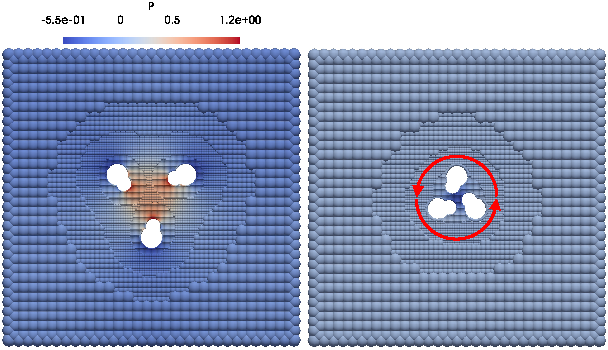}}
\caption{Three colloidal dimers initially lying apart (\textit{left}) assembling into a chiral cluster (\textit{right}). Depending on their initial relative orientations, they form either a (a) left-handed or (b) right-handed chiral cluster. The equilibrated chiral cluster of three dimers rotates (a) clockwise or (b) counter-clockwise with a steady rotational speed. The snapshots also depict the distributions of GMLS particles with adaptive refinement, with an average of 3350 GMLS particles used. The color is correlated to the computed fluid pressure (Unit in color bar: $10^{-9}~\si{Pa})$.}
\label{fig:3colloid-pressure}
\end{figure}

\begin{figure}[!htp]
\centering
\subfigure[]{
\includegraphics[width=5.0in]{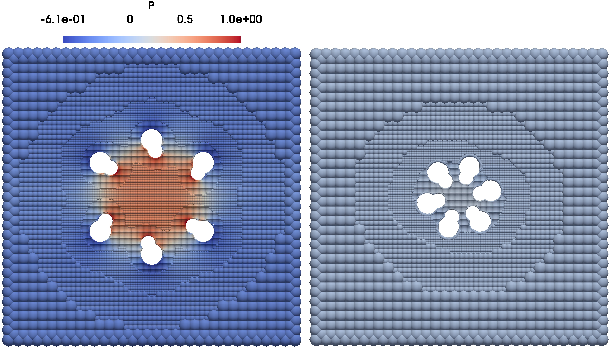}}
\subfigure[]{
\includegraphics[width=5.0in]{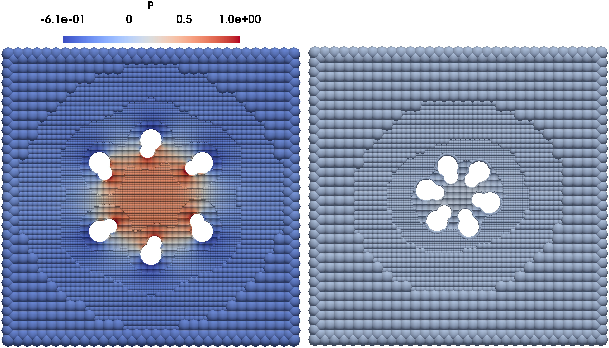}}
\caption{Six colloidal dimers initially lying apart (\textit{left}) assembling into an achiral cluster (\textit{right}), independent of their initial relative orientations. The equilibrated achiral clusters of six dimers rest without rotating. The snapshots also depict the distributions of GMLS particles with adaptive refinement, with an average of 5870 GMLS particles used. The color is correlated to the computed fluid pressure (Unit in color bar: $10^{-9}~\si{Pa})$.}
\label{fig:6colloid-pressure}
\end{figure}

Lastly, we solved a more complicated but interesting problem, regarding the collective dynamics of multiple active asymmetric colloidal dimers. This study was motivated by an experimental work \cite{ma2015electric}, where each dimer was composed of two overlapped spheres of different radii. As the dynamics of interest for these dimers occurred only along the substrate plane in the experiment, it may be posed as a 2D problem. Thus, each dimer was represented by two overlapped circular colloids in our simulations. While the larger lobe of each dimer had a radius of $R_L = 4\times10^{-6}~\si{m}$, the radius of the smaller lobe was: $R_S = 2.8\times10^{-6}~\si{m}$. The center-to-center distance of each dimer's two lobes was equal to $R_L$. A collection of such colloidal dimers were suspended in a square domain with each edge length of $L = 100\times10^{-6}~\si{m}$. The snapshots from the simulations shown in Figures \ref{fig:3colloid-pressure} and \ref{fig:6colloid-pressure} also illustrate the configuration of this problem. The fluid was assumed to be water with $\rho = 1.0\times 10^3~\si{kg/m^3}$ and $\nu = 1.0\times 10^{-6}~\si{m^2/s}$. Due to the small length scale and slow dynamics \cite{ma2015electric}, this may be treated as a Stokes problem. 

According to the experimental work \cite{ma2015electric}, each colloidal dimer can lie on the substrate surface and be actively moving in the fluid environment, propelled by an external field. The active motion of each dimer is along the center line of the two lobes toward the smaller lobe. As any two dimers come sufficiently close, a repulsive force acts between the two larger lobes and also between the two smaller lobes of the two dimers. Due to the presence of a standing, stationary dimer, the active, lying dimers nearby can be attracted toward it and form clusters of 2 to 6 dimers. It is the interplay and competition between the aforementioned active motion, repulsive force, and attraction that determine the collective dynamics and clustering of the dimers. In our simulations, the active motion of each dimer was driven by a constant force $F_e = 1\times10^{-9}~\si{N}$ exerted along the center line of its two lobes toward the smaller lobe. The repulsive force between lobes was modeled by: $F_r = R^2(\frac{2R}{r})^4$, where $R = R_L$ or $R_S$, and $r$ is the center-to-center distance of the two interacting lobes. This repulsive force was applied along the center line between the two larger lobes or between the two smaller lobes of two dimers. Note that a torque could be generated for each dimer by this repulsive force. The attraction between dimers was represented by a force $F_a = R^2(\frac{R}{r})^2$, where $R = R_L$ or $R_S$, and $r$ is the distance from the center of the larger/smaller lobe to the standing dimer. The force was exerted at the centers of the two lobes of each dimer and toward the position of the standing dimer. A torque could also be generated for each dimer by this attractive force.
%

Using the adaptive GMLS, we simulated two groups of dimers: the first with three dimers and the second with six dimers. In Figure \ref{fig:3colloid-pressure}, three dimers initially lying apart may form a chiral cluster. The chirality (left-handedness or right-handedness) depends on the initial relative orientations of the three dimers. The equilibrated chiral cluster of three dimers rotates clockwise or counter-clockwise, depending on its chirality, with a steady rotational speed. (Refer to the animation in the Supporting Information.) In Figure \ref{fig:6colloid-pressure}, six dimers initially set apart finally assemble into an achiral cluster, regardless of their initial relative orientations. The equilibrated achiral clusters of six dimers remain stationary without rotating or moving. (Refer to the animation in the Supporting Information.) These different collective and clustering behaviors are determined by the competition between the active motions of dimers and the repulsive and attractive forces between the dimers. Our predictions for both groups of dimers are consistent with the experimental observations \cite{ma2015electric}. We hence demonstrate that the proposed adaptive GMLS is applicable to address more complicated but practical applications. 
%

%% file: conclusion.tex
\section{Conclusion} \label{sec:Conclu}

We have introduced a spatially adaptive high-order meshless method. It employs the GMLS discretization for solving the governing PDEs, and hence inherits the GMLS' meshless nature and attributes of consistency, compatibility, and high-order accuracy. By applying the recovery-based  \textit{a posteriori} error estimator as the adaptive criterion, the method introduces spatial adaptivity in a rigorous and optimal way. Following the four-step adaptive algorithm, the spatial resolution can be adaptively refined near point singularities and fluid-solid interfaces. The method is found flexible and robust in solving various div-grad and Stokes problems with corner singularity, discontinuous diffusion coefficient, and moving boundaries of non-trivial geometries. The proposed recovery-based error displayed the same convergence order as the true errors, providing accurate estimations of the true errors. The numerical solutions produced by the adaptive GMLS method could achieve optimal convergence orders even in the presence of singularities, matching the convergence for non-singular solutions and regular domains in the literature \cite{Trask2017high,Trask2018compatible}. Compared with the GMLS solutions with uniform resolution, the adaptive GMLS solutions yield the same accuracy but used much fewer degrees of freedom. In the simulations of colloidal suspensions, by employing adaptive refinement in lubrication gaps and sharp representation of fluid-solid interfaces of arbitrary geometries, the short-range lubrication interaction between colloids was resolved accurately and efficiently without invoking any artificial subgrid-scale lubrication model. Thus, it is anticipated that the proposed adaptive GMLS method can be applied for solving inherently multiscale problems. 

Looking ahead, the adaptive GMLS can be extended for solving FSI in a wide range of Reynolds numbers, beyond the Stokes problems. Its meshless nature will allow for advecting the GMLS particles with the flow and hence solving the Navier-Stokes equation in a Lagrangian sense. In addition, although the simulations herein were conducted in 2D, the proposed method is directly applicable to a 3D implementation, which insofar efficiency gains are concerned, stands to benefit even more from the proposed adaptive refinement strategy. With these extensions, we expect that the proposed spatially adaptive high-order meshless method can be applied to address a multitude of FSI applications.

%% file: main.bbl
\begin{thebibliography}{10}
\expandafter\ifx\csname url\endcsname\relax
  \def\url#1{\texttt{#1}}\fi
\expandafter\ifx\csname urlprefix\endcsname\relax\def\urlprefix{URL }\fi
\expandafter\ifx\csname href\endcsname\relax
  \def\href#1#2{#2} \def\path#1{#1}\fi

\bibitem{ActiveColloid_ReviewAlexeev2016}
S.~V. Nikolov, H.~Shum, A.~C. Balazs, A.~Alexeev, Computational design of
  microscopic swimmers and capsules: From directed motion to collective
  behavior, Current Opinion in Colloid \& Interface Science 21 (2016) 44--56.

\bibitem{ActiveColloid_Review2016b}
A.~E. Patteson, A.~Gopinath, P.~E. Arratia, Active colloids in complex fluids,
  Current Opinion in Colloid \& Interface Science 21 (2016) 86--96.

\bibitem{ma2015electric}
F.~Ma, S.~Wang, D.~T. Wu, N.~Wu, Electric-field--induced assembly and
  propulsion of chiral colloidal clusters, Proceedings of the National Academy
  of Sciences (2015) 201502141.

\bibitem{bueno2015interaction}
J.~Bueno, C.~Bona-Casas, Y.~Bazilevs, H.~Gomez, Interaction of complex fluids
  and solids: theory, algorithms and application to phase-change-driven
  implosion, Computational Mechanics 55~(6) (2015) 1105--1118.

\bibitem{ten2014gravitaxis}
B.~Ten~Hagen, F.~K{\"u}mmel, R.~Wittkowski, D.~Takagi, H.~L{\"o}wen,
  C.~Bechinger, Gravitaxis of asymmetric self-propelled colloidal particles,
  Nature communications 5 (2014) 4829.

\bibitem{goto2015purely}
Y.~Goto, H.~Tanaka, Purely hydrodynamic ordering of rotating disks at a finite
  reynolds number, Nature Communications 6 (2015) 5994.

\bibitem{randles2017computational}
A.~Randles, D.~H. Frakes, J.~A. Leopold, Computational fluid dynamics and
  additive manufacturing to diagnose and treat cardiovascular disease, Trends
  in Biotechnology 35~(11) (2017) 1049--1061.

\bibitem{kamps2018design}
T.~Kamps, M.~Biedermann, C.~Seidel, G.~Reinhart, Design approach for additive
  manufacturing employing constructal theory for point-to-circle flows,
  Additive Manufacturing 20 (2018) 111--118.

\bibitem{sochol20163d}
R.~Sochol, E.~Sweet, C.~Glick, S.~Venkatesh, A.~Avetisyan, K.~Ekman,
  A.~Raulinaitis, A.~Tsai, A.~Wienkers, K.~Korner, et~al., 3d printed
  microfluidic circuitry via multijet-based additive manufacturing, Lab on a
  Chip 16~(4) (2016) 668--678.

\bibitem{WindTurbineFSIIGA2012}
Y.~Bazilevs, M.-C. Hsu, M.~Scott, Isogeometric fluid--structure interaction
  analysis with emphasis on non-matching discretizations, and with application
  to wind turbines, Computer Methods in Applied Mechanics and Engineering
  249-252 (2012) 28--41.

\bibitem{EnergyWaveFSI2008}
E.~B. Agamloh, A.~K. Wallace, A.~von Jouanne, Application of fluid--structure
  interaction simulation of an ocean wave energy extraction device, Renewable
  Energy 33~(4) (2008) 748--757.

\bibitem{WindTurbineFSIIGA2016}
J.~Yan, A.~Korobenko, X.~Deng, Y.~Bazilevs, Computational free-surface
  fluid--structure interaction with application to floating offshore wind
  turbines, Computers \& Fluids 141 (2016) 155--174.

\bibitem{EnergyWaveFSI2018}
A.~Calderer, X.~Guo, L.~Shen, F.~Sotiropoulos, Fluid--structure interaction
  simulation of floating structures interacting with complex, large-scale ocean
  waves and atmospheric turbulence with application to floating offshore wind
  turbines, Journal of Computational Physics 355 (2018) 144--175.

\bibitem{Trask2017high}
N.~Trask, M.~Perego, P.~Bochev, A high-order staggered meshless method for
  elliptic problems, SIAM Journal on Scientific Computing 39~(2) (2017)
  A479--A502.

\bibitem{Trask2018compatible}
N.~Trask, M.~Maxey, X.~Hu, A compatible high-order meshless method for the
  stokes equations with applications to suspension flows, Journal of
  Computational Physics 355 (2018) 310--326.

\bibitem{wendland2004scattered}
H.~Wendland, Scattered data approximation, Vol.~17, Cambridge University Press,
  2004.

\bibitem{mirzaei2012generalized}
D.~Mirzaei, R.~Schaback, M.~Dehghan, On generalized moving least squares and
  diffuse derivatives, IMA Journal of Numerical Analysis 32~(3) (2012)
  983--1000.

\bibitem{Pan_LLNS_2014}
J.~Kordilla, W.~Pan, A.~Tartakovsky, Smoothed particle hydrodynamics model for
  {Landau-Lifshitz-Navier-Stokes} and advection-diffusion equations, The
  Journal of Chemical Physics 141~(22) (2014) 224112.

\bibitem{Pan_CBFSPH_2014}
W.~Pan, J.~Bao, A.~M. Tartakovsky, Smoothed particle hydrodynamics continuous
  boundary force method for {N}avier--{S}tokes equations subject to a {R}obin
  boundary condition, Journal of Computational Physics 259 (2014) 242--259.

\bibitem{Pan_ISPH2_2017}
W.~Pan, K.~Kim, M.~Perego, A.~M. Tartakovsky, M.~L. Parks, Modeling
  electrokinetic flows by consistent implicit incompressible smoothed particle
  hydrodynamics, Journal of Computational Physics 334 (2017) 125--144.

\bibitem{Trask2015SPH}
N.~Trask, M.~Maxey, K.~Kimb, M.~Perego, M.~L. Parks, K.~Yang, J.~Xu, A scalable
  consistent second-order {SPH} solver for unsteady low {R}eynolds number
  flows, Computer Methods in Applied Mechanics and Engineering 289 (2015)
  155--178.

\bibitem{Pan_CMRSPH_2017}
W.~Hu, W.~Pan, M.~Rakhsha, Q.~Tian, H.~Hu, D.~Negrut, A consistent
  multi-resolution smoothed particle hydrodynamics method, Computer Methods in
  Applied Mechanics and Engineering 324 (2017) 278--299.

\bibitem{Pan_ARSPH_2019}
W.~Hu, G.~Guo, X.~Hu, D.~Negrut, Z.~Xu, W.~Pan, A consistent spatially adaptive
  smoothed particle hydrodynamics method for fluid-structure interactions,
  Computer Methods in Applied Mechanics and Engineering 347 (2019) 402--424.

\bibitem{VerfurthR1996a}
R.~Verf{\"u}rth, A review of a posteriori error estimation and adaptive
  mesh-refinement techniques, John Wiley \& Sons, 1996.

\bibitem{nochetto2009theory}
R.~H. Nochetto, K.~G. Siebert, A.~Veeser, Theory of adaptive finite element
  methods: an introduction, in: Multiscale, nonlinear and adaptive
  approximation, Springer, 2009, pp. 409--542.

\bibitem{Zienkiewicz1992}
O.~Zienkiewicz, J.~Zhu, The superconvergent patch recovery {SPR} and adaptive
  finite element refinement, Computer Methods in Applied Mechanics and
  Engineering 101~(1-3) (1992) 207--224.

\bibitem{RecoverError_Book2000}
M.~Ainsworth, J.~T. Oden, A Posteriori Error Estimation in Finite Element
  Analysis, John Wiley \& Sons, 2000, Ch. Recovery-Based Error Estimators, pp.
  65--84.

\bibitem{Xu.J;Zhang.Z2004}
J.~Xu, Z.~Zhang, Analysis of recovery type a posteriori error estimators for
  mildly structured grids, Mathematics of Computation 73~(247) (2004)
  1139--1152.

\bibitem{RecoverError_Zhang2005}
Z.~Zhang, A.~Naga, A new finite element gradient recovery method:
  {S}uperconvergence property, SIAM Journal on Scientific Computing 26~(4)
  (2005) 1192--1213.

\bibitem{RecoverError_Zhang2004}
A.~Naga, Z.~Zhang, A posteriori error estimates based on the polynomial
  preserving recovery, SIAM Journal on Numerical Analysis 42~(4) (2004)
  1780--1800.

\bibitem{dorfler1996convergent}
W.~D{\"o}rfler, A convergent adaptive algorithm for poisson’s equation, SIAM
  Journal on Numerical Analysis 33~(3) (1996) 1106--1124.

\bibitem{stevenson2007optimality}
R.~Stevenson, Optimality of a standard adaptive finite element method,
  Foundations of Computational Mathematics 7~(2) (2007) 245--269.

\bibitem{butcher2016numerical}
J.~C. Butcher, Numerical methods for ordinary differential equations, John
  Wiley \& Sons, 2016.

\bibitem{CompatibleDis_Book2006}
P.~B. Bochev, J.~M. Hyman, Principles of mimetic discretizations of
  differential operators, in: Compatible spatial discretizations, Springer,
  2006, pp. 89--119.

\bibitem{MimeticFD_JCP2014}
K.~Lipnikov, G.~Manzini, M.~Shashkov, Mimetic finite difference method, Journal
  of Computational Physics 257 (2014) 1163--1227.

\bibitem{ainsworth2000posteriori}
M.~Ainsworth, J.~T. Oden, A Posteriori Error Estimation in Finite Element
  Analysis, John Wiley \& Sons, 2000.

\bibitem{Saad2003a}
Y.~Saad, Iterative Methods for Sparse Linear Systems, 2nd Edition, {Society for
  Industrial and Applied Mathematics}, Philadelphia, PA, 2003.

\bibitem{schiff1988finite}
B.~Schiff, Finite element eigenvalues for the {L}aplacian over an {L}-shaped
  domain, Journal of Computational Physics 76~(2) (1988) 233--242.

\bibitem{johannessen2014isogeometric}
K.~A. Johannessen, T.~Kvamsdal, T.~Dokken, Isogeometric analysis using {LR}
  {B}-splines, Computer Methods in Applied Mechanics and Engineering 269 (2014)
  471--514.

\bibitem{DGPorousMedia2010}
A.~Ern, I.~Mozolevski, L.~Schuh, Discontinuous {G}alerkin approximation of
  two-phase flows in heterogeneous porous media with discontinuous capillary
  pressures, Computer Methods in Applied Mechanics and Engineering 199~(23)
  (2010) 1491--1501.

\bibitem{MultiscaleFEMPorousMediaHOU1997}
T.~Y. Hou, X.-H. Wu, A multiscale finite element method for elliptic problems
  in composite materials and porous media, Journal of Computational Physics
  134~(1) (1997) 169--189.

\bibitem{DGPorousMedia1998}
R.~Helmig, R.~Huber, Comparison of {G}alerkin-type discretization techniques
  for two-phase flow in heterogeneous porous media, Advances in Water Resources
  21~(8) (1998) 697--711.

\bibitem{PorousMedia1991}
L.~J. Durlofsky, Numerical calculation of equivalent grid block permeability
  tensors for heterogeneous porous media, Water Resources Research 27~(5)
  (1991) 699--708.

\bibitem{Electrostatics2013}
V.~Jadhao, F.~J. Solis, M.~Olvera de~la Cruz, A variational formulation of
  electrostatics in a medium with spatially varying dielectric permittivity,
  The Journal of Chemical Physics 138~(5) (2013) 054119.

\bibitem{luo2010modeling}
X.~Luo, A.~Beskok, G.~E. Karniadakis, Modeling electrokinetic flows by the
  smoothed profile method, Journal of Computational Physics 229~(10) (2010)
  3828--3847.

\bibitem{Electrostatics1991}
M.~E. Davis, J.~A. McCammon, Dielectric boundary smoothing in finite difference
  solutions of the {P}oisson equation: {A}n approach to improve accuracy and
  convergence, Journal of Computational Chemistry 12~(7) (1991) 909--912.

\bibitem{epperson1987runge}
J.~F. Epperson, On the {R}unge example, The American Mathematical Monthly
  94~(4) (1987) 329--341.

\bibitem{michael2002scientific}
T.~H. Michael, Scientific computing: an introductory survey, The McGraw-540
  Hill Companies Inc.: New York, NY, USA.

\bibitem{wannier1950contribution}
G.~H. Wannier, A contribution to the hydrodynamics of lubrication, Quarterly of
  Applied Mathematics 8~(1) (1950) 1--32.

\bibitem{yuan1994rheology}
X.~Yuan, R.~Ball, Rheology of hydrodynamically interacting concentrated hard
  disks, The Journal of Chemical Physics 101~(10) (1994) 9016--9021.

\bibitem{kromkamp2006lattice}
J.~Kromkamp, D.~van~den Ende, D.~Kandhai, R.~van~der Sman, R.~Boom, Lattice
  {B}oltzmann simulation of {2D} and {3D} non-{B}rownian suspensions in
  {C}ouette flow, Chemical Engineering Science 61~(2) (2006) 858--873.

\bibitem{YEOJCP2010}
K.~Yeo, M.~R. Maxey, Simulation of concentrated suspensions using the
  force-coupling method, Journal of Computational Physics 229~(6) (2010)
  2401--2421.

\bibitem{darabaner1967particle}
C.~Darabaner, J.~Raasch, S.~Mason, Particle motions in sheared suspensions
  {XX}: {C}ircular cylinders, The Canadian Journal of Chemical Engineering
  45~(1) (1967) 3--12.

\bibitem{bian2014splitting}
X.~Bian, M.~Ellero, A splitting integration scheme for the {SPH} simulation of
  concentrated particle suspensions, Computer Physics Communications 185~(1)
  (2014) 53--62.

\end{thebibliography}
